\documentclass[fleqn,usenatbib]{mnras}

\usepackage[T1]{fontenc}

\DeclareRobustCommand{\VAN}[3]{#2}
\let\VANthebibliography\thebibliography
\def\thebibliography{\DeclareRobustCommand{\VAN}[3]{##3}\VANthebibliography}

\usepackage{graphicx}	
\usepackage{amsmath}	
\usepackage{bm}	        
\usepackage{enumitem}
\usepackage{csvsimple}
\usepackage{multirow}
\usepackage{siunitx} 
\usepackage[utf8]{inputenc}
\usepackage{dirtytalk}
\usepackage{xparse}

\makeatletter
\newlength{\abovecaptionskip}%
\makeatother
\usepackage[para,online,flushleft]{threeparttable}
\usepackage{txfonts}
\usepackage[pdfpagelabels=false]{hyperref}	
\hypersetup{colorlinks=true,linkcolor=blue,citecolor=blue,filecolor=blue,urlcolor=blue,}

\title[Low-$J$ CO excitation across NGC\,2903 \& NGC\,3627]{Resolved low-$J$ $^{12}$CO excitation at $190$~parsec resolution across NGC\,2903 and NGC\,3627}

\author[den Brok et al.]{
J. S. den Brok,$^{1}$\thanks{E-mail: jakob.denbrok@gmail.com},
A. K. Leroy$^{2}$,
A. Usero$^{3}$,
E. Schinnerer$^{4}$,
E. Rosolowsky$^{5}$,
E. W. Koch$^{1}$, \newauthor
M. Querejeta$^{3}$, 
D. Liu$^{6}$,
F. Bigiel$^{7}$,
A. T. Barnes$^{8}$,
M. Chevance$^{9,10}$,
D. Colombo$^{7}$,
D. A. Dale$^{11}$, \newauthor
S.~C.~O. Glover$^{9}$,
M. J. Jimenez-Donaire$^{3,12}$,
Y.-H. Teng$^{13}$,
T. G. Williams$^{14}$\\
$^{1}$Center for Astrophysics $\mid$ Harvard \& Smithsonian, 60 Garden St., 02138 Cambridge, MA, USA\\
$^{2}$Department of Astronomy, The Ohio State University, 140 West 18th Avenue, Columbus, OH 43210\\
$^{3}$Observatorio Astronómico Nacional (IGN), C/ Alfonso XII, 3, 28014, Madrid, Spain\\
$^{4}$Max Planck Institute for Astronomy, Königstuhl 17, D-69117, Germany\\
$^{5}$Department of Physics, University of Alberta, Edmonton, AB T6G 2E1, Canada\\
$^{6}$Max-Planck-Institut für Extraterrestrische Physik (MPE), Giessenbachstr. 1, D-85748 Garching, Germany\\
$^{7}$Argelander-Institut für Astronomie, Universität Bonn, Auf dem Hügel 71, D-53121 Bonn, Germany\\
$^{8}$European Southern Observatory, Karl-Schwarzschild Straße 2, 85748, Garching bei München, Germany\\
$^{9}$Universität Heidelberg, Zentrum für Astronomie, Institut für Theoretische Astrophysik, Albert-Ueberle-Straße 2, D-69120 Heidelberg, Germany\\
$^{10}$ Cosmic Origins Of Life (COOL) Research DAO\\
$^{11}$Department of Physics and Astronomy, University of Wyoming, Laramie, WY 82071, USA\\
$^{12}$Centro de Desarrollos Tecnológicos, Observatorio de Yebes (IGN), 19141 Yebes, Guadalajara, Spain\\
$^{13}$Center for Astrophysics and Space Sciences, Department of Physics, University of California San Diego, 9500 Gilman Dr., La Jolla, CA 92093, USA\\
$^{14}$Sub-department of Astrophysics, Department of Physics, University of Oxford, Keble Road, Oxford OX1 3RH, UK
}

\date{Accepted 2023 October 3. Received 2023 October 1; in original form 2023 August 2}

\pubyear{2023}

\begin{document}

\renewcommand{\figureautorefname}{Fig.} 
\renewcommand{\equationautorefname}{Eq.} 
\renewcommand{\sectionautorefname}{Section} 
\renewcommand{\subsectionautorefname}{Section}
\renewcommand{\subsubsectionautorefname}{Section}
\renewcommand{\appendixautorefname}{Appendix}





\newcommand{  \Hi      }{\ifmmode {\rm H}\,\textsc{i} \else H\,\textsc{i}\fi}

\newcommand*{\species}[1]{%
    \ensuremath{\IfBold{\bm{\mathrm{#1}}}{\mathrm{#1}}}%
}

\newcommand*{\trans}[1]{%
    \ensuremath{\IfBold{\bm{\mytrans{#1}}}{\mytrans{#1}}}%
}

\newcommand{\delim}{{-}}
\newcommand*{\mytrans}[1]{%
    \ifnum#1=10 (1\delim0)\else%
    \ifnum#1=21 (2\delim1)\else%
    \ifnum#1=32 (3\delim2)\else%
    \ifnum#1=43 (4\delim3)\else%
    \ifnum#1=54 (5\delim4)\else%
    \ifnum#1=65 (6\delim5)\else%
    \ifnum#1=76 (7\delim6)\else%
    \ifnum#1=87 (8\delim7)\else%
    \ifnum#1=98 (9\delim8)\else%
    \ifnum#1=109 (10\delim9)\else#1%
    \fi\fi\fi\fi\fi\fi\fi\fi\fi\fi%
}

\makeatletter
\newcommand*{\IfBold}{%
  \ifx\f@series\my@test@bx
    \expandafter\@firstoftwo
  \else
    \expandafter\@secondoftwo
  \fi
}
\newcommand*{\my@test@bx}{bx}
\makeatother

\DeclareDocumentCommand{\chem}{ m g }{%
    {\species{#1}%
        \IfNoValueF {#2} {\,\trans{#2}}%
    }%
}

\label{firstpage}
\pagerange{\pageref{firstpage}--\pageref{lastpage}}
\maketitle

\begin{abstract}
The low-$J$ rotational transitions of \chem{^{12}CO} are commonly used to trace the distribution of molecular gas in galaxies. Their ratios are sensitive to excitation and physical conditions in the molecular gas. Spatially resolved studies of CO ratios are still sparse and affected by flux calibration uncertainties, especially since most do not have high angular resolution or do not have short-spacing information and hence miss any diffuse emission. We compare the low-$J$ CO ratios across the disk of two massive, star-forming spiral galaxies NGC\,2903 and NGC\,3627 to investigate whether and how local environments drive excitation variations at GMC scales. We use Atacama Large Millimeter Array (ALMA) observations of the three lowest-$J$ CO transitions at a common angular resolution of 4$''$ (190\,pc). 
We measure median line ratios of $R_{21}=0.67^{+0.13}_{-0.11}$, $R_{32}=0.33^{+0.09}_{-0.08}$, and $R_{31}=0.24^{+0.10}_{-0.09}$ across the full disk of NGC\,3627. We see clear CO line ratio variation across the galaxy consistent with changes in temperature and density of the molecular gas. In particular, toward the center, $R_{21}$, $R_{32}$, and $R_{31}$ increase by 35\%, 50\%, and 66\%, respectively compared to their average disk values. 
The overall line ratio trends suggest that CO(3-2) is more sensitive to changes in the excitation conditions than the two lower-$J$ transitions. Furthermore, we find a similar radial $R_{32}$ trend in NGC\,2903, albite a larger disk-wide average of $\langle R_{32}\rangle=0.47^{+0.14}_{-0.08}$. We conclude that the CO low-$J$ line ratios vary across environments in such a way that they can trace changes in the molecular gas conditions, with the main driver being changes in temperature. 
\end{abstract}




\section{Introduction}

Emission by rotational transitions of carbon monoxide (CO) constitutes the primary tracer of the bulk molecular gas mass distribution and gas kinematics in extragalactic studies. Over the past decades, various rotational $J$ transitions of CO have been targeted and surveyed in and across ${>}1{,}000$ galaxies at low and high $z$ \citep[e.g.,][]{Carilli2013, Saintonge2017, Liu2019a,Liu2019b}. Predominantly, such studies have observed CO in a galaxy-wide, integrated fashion or at coarse, kpc-scale spatial resolution. In recent years, however, the number of nearby galaxies that have been mapped {{down to}} giant molecular cloud (GMC)  scales has increased to ${\sim}100$ \citep[e.g.,][]{DonovanMeyer2012, Schinnerer2013,Leroy2021, Koda2023}. 
{{The molecular gas properties of GMCs are affected by their local environment \citep{Sun2020}. Environments in galaxies that show more extreme molecular gas conditions  (including higher temperature, molecular gas density, and star formation rate), are the center and bar-ends, which have sizes of ${\sim}1{-}2\,$kpc \citep{Law2018,Sun2020, denbrok2021}. At sub-kpc physical resolution, it is therefore possible to resolve the different environments into several independent sightlines. }}
%

Historically, the focus has been on observing the low-$J$ CO(1-0) transition at a rest-frequency of $\nu{\approx}115\,\rm GHz$. But recent improvements in sub-mm facilities have made it possible to map nearby galaxies using higher rest-frequency lines, such as CO(2-1) and CO(3-2), which allow for higher-resolution observations \citep[e.g.,][]{Leroy2021}. In addition, high-$z$ studies regularly observe even higher-$J$ rotational lines, which are redshifted into the mm regime \citep[e.g.,][]{Tacconi2013,Dessauges2017, Daddi2015, Harrington2021,Liu2021,Lenkic2023}. 
    
    
%
%

\begin{figure*}
    \centering
    \includegraphics[width = \textwidth]{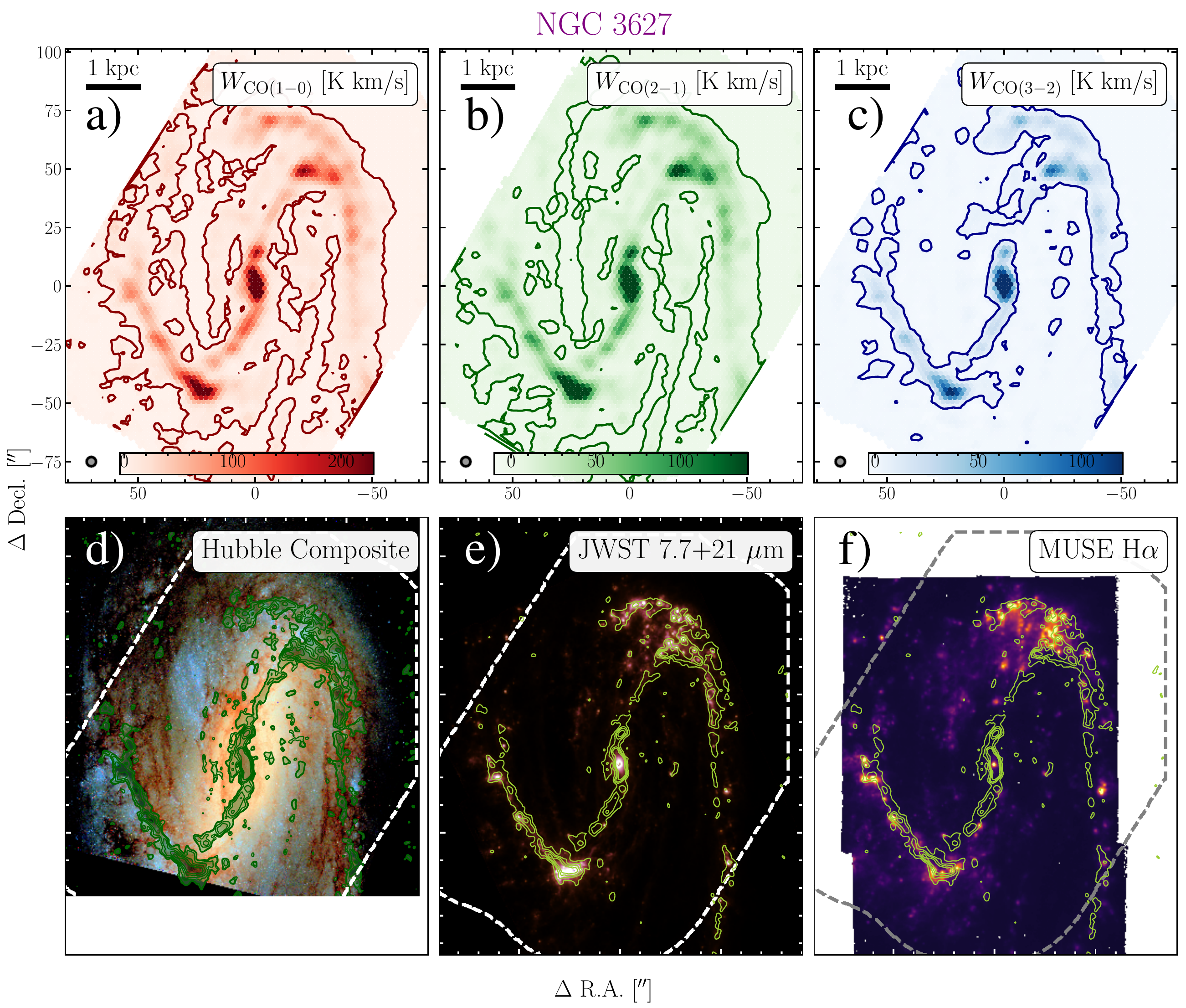}
    \caption{{\bf CO moment-0 maps of a large part of the disk of NGC 3627} The three top panels (a--c) present the integrated intensity maps (moment-0) of \chem{^{12}CO}{10}, \trans{21}, and \trans{32} respectively. These three maps have been convolved to a common beam size of 4$''$ (indicated by the circle in the lower left corner). The thicker colored line indicates the $\rm S/N{>}5$ threshold. The pixels are half-beam size spaced in a hexagonal grid. The bottom three panels (d--f) show ancillary data sets: A composite Hubble Space Telescope Image (credit: ESA/Hubble, NASA) of the same field-of-view (the white-dashed line indicates the exact field-of-view of the CO(1-0) observations) with green contours showing the \chem{^{12}CO}{21} integrated intensity at arbitrary intervals for reference purposes (d); a color composite of the  JWST 7.7 and 21 $\mu$m band tracing dust (e) and  the MUSE extinction-corrected H$\alpha$ image (f).}
    \label{fig:ngc3627_map}
\end{figure*}

\begin{figure*}
    \centering
    \includegraphics[width = \textwidth]{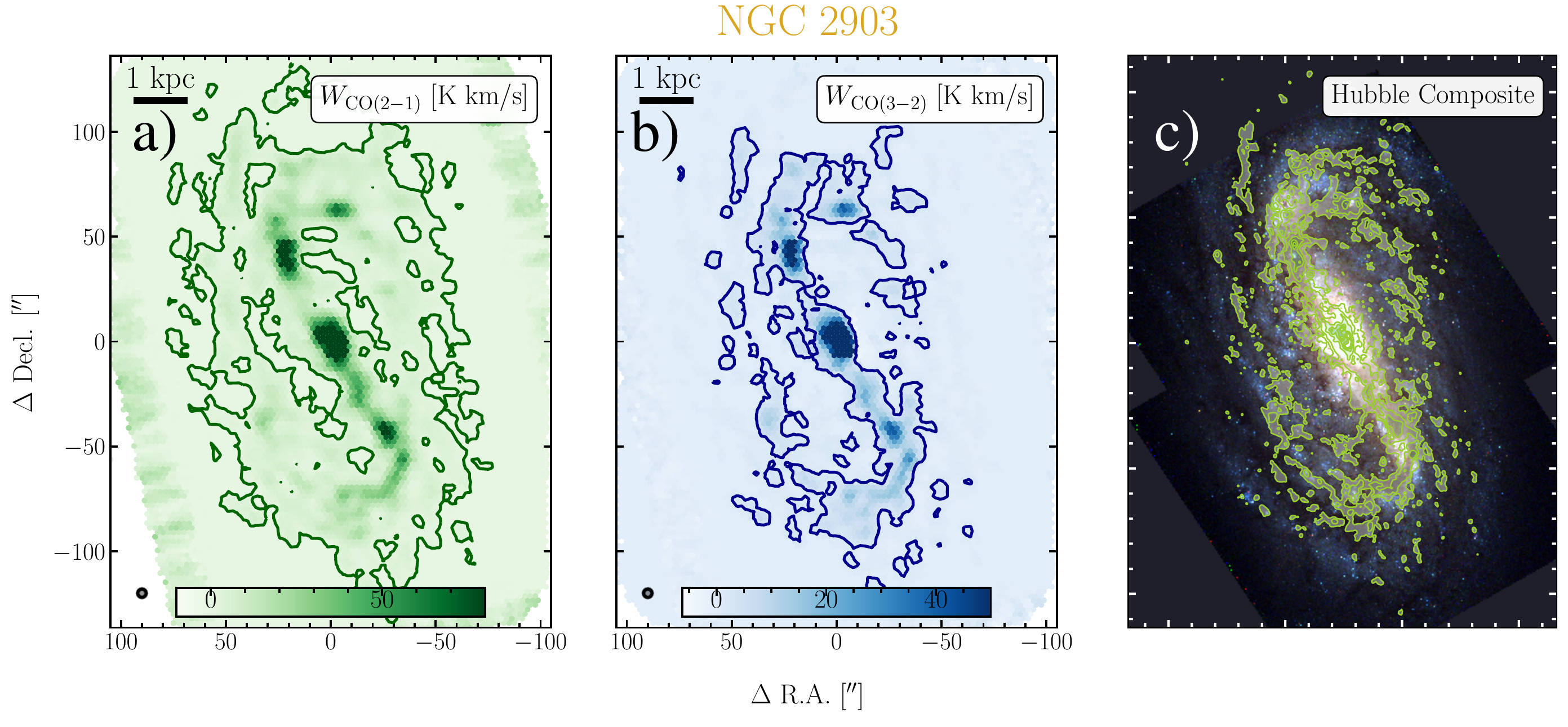}
    \caption{{\bf CO moment-0 maps of the entire disk of NGC 2903} Panels a) and b) illustrate the integrated intensities (moment-0) of \trans{21} and \trans{32}, respectively. These maps have been convolved to a common beam size of 4$''$ (indicated by the circle in the lower left corner). The description follows \autoref{fig:ngc3627_map}. Panel c) depicts a composite Hubble Space Telescope Image (credit: ESA/Hubble, NASA, and L. Ho, J. Lee and the PHANGS-HST Team). Unlike for NGC\,3627, JWST or MUSE observations are not yet available.}
    \label{fig:ngc2903_map}
\end{figure*}

Some of the main uncertainties that limit our understanding of molecular gas are interpreting CO emission and gauging its connection to the physical state of the gas. Multi-transition molecular spectroscopy represents the best path to do this, offering a method to probe the excitation, optical depth, density, and temperature of the gas.
Since different studies target different CO rotational transitions, comparing and contrasting the results is not straightforward. The reason for this is that the different CO line intensity ratios are sensitive to changes in the physical conditions of the molecular gas, such as its temperature, column density, and opacity \citep[e.g., ][]{Penaloza2017}. 
For instance, CO(1-0) has an optically thin critical density of $n_{\rm crit.}\approx 2\times10^3{\rm cm^{-3}}$, CO(2-1) has $n_{\rm crit.}\approx 1\times10^4{\rm cm^{-3}}$, and  CO(3-2) has $n_{\rm crit.}\approx 4\times10^4{\rm cm^{-3}}$ \citep{Carilli2013}. Since the CO emission is generally optically thick, the effective critical densities in real GMCs are even lower \citep{Shirley2015}.  In turn, using several rotational-$J$ lines is useful to constrain the properties of the gas, but previous studies are generally limited to low angular (kpc-scale) resolution \citep[e.g.][]{Greve2014,Liu2015,Rosenberg2015}, or only one or two lines at high angular resolution.

Several previous studies have focused on assessing low-$J$ CO line ratio variations across a sample of nearby galaxies \citep[e.g.,][]{Crosthwaite2007,Wilson2012, denbrok2021,Egusa2022, Leroy2022}. At kpc-scale resolution, significant variation in the brightness temperature ratios $R_{21}{\equiv}$CO(2-1)/(1-0) and $R_{32}{\equiv}$CO(3-2)/(2-1) with negative radial and positive correlation with the star formation rate surface density has been detected \citep[e.g.,][]{denbrok2021,Yajima2021,Leroy2022}. 
However, it remains unclear how CO line ratios vary from region to region within a galaxy, i.e., on {sub-kpc scales.} 
Moreover, while high resolution, short-spacing corrected ALMA observations of CO(2-1) and (1-0) are becoming more common (e.g. T. Saito et al. in prep.), similar mapping of the CO(3-2) line remains rare. Most of our resolved maps of this line still come from array receivers on single dish telescopes \citep[e.g.,][]{Wilson2012} with resolutions of ${\sim}10''{-}20''$.

In this paper we report new ALMA CO~(3-2) mapping of two nearby star-forming galaxies, NGC 3627 and NGC 2903. They both are star-forming spiral galaxies at a similar distance ($D_{3627}{=}8.8$\,Mpc and $D_{2903}{=}9.3$\,Mpc; \citealt{Anand2021}) and both have a prominent bar. However, while NGC\,3627 is interacting, NGC\,2903 is an isolated galaxy. Furthermore, NGC\,3627 hosts a low-luminous Seyfert 2 AGN, but NGC\,2903 is inactive. Combined with previous ALMA CO~(2-1) data, this new imaging allows us to measure the $R_{32}$ line ratio at $4''$ (< 200~pc) resolution in both galaxies. In NGC\,3627 we also use archival ALMA CO~(1-0) mapping, and so measure $R_{21}$, $R_{32}$, and $R_{31}$ at 4$''$ (${\sim}$190pc) resolution across the whole galaxy. 
We address two topics in this study:
\begin{enumerate}
    \item At 190 pc resolution, do the line ratios $R_{21}$, $R_{32}$, and $R_{31}$ track one another? How localized are their variations to specific environments, such as the galaxy center and bar ends? And how much local variation in the ratio do we observe at this high resolution compared to previous $\sim$kpc resolution studies?
    \item 
    What are the implications of the observed line ratio variations for changes in the temperature and density of the molecular gas?
\end{enumerate}

This study is structured as follows: In \autoref{sec:data}, we provide a brief overview of the selected observations. \autoref{sec:result} describes the line ratio variation across different environments and angular scales. In \autoref{sec:discussion}, we discuss how the line ratios probe the molecular gas conditions throughout NGC\,3627, and in \autoref{sec:conclusion}, we summarize and conclude on our findings.


\section{Data}
\label{sec:data}

\begin{figure*}
    \centering
    \includegraphics[width =0.85\textwidth]{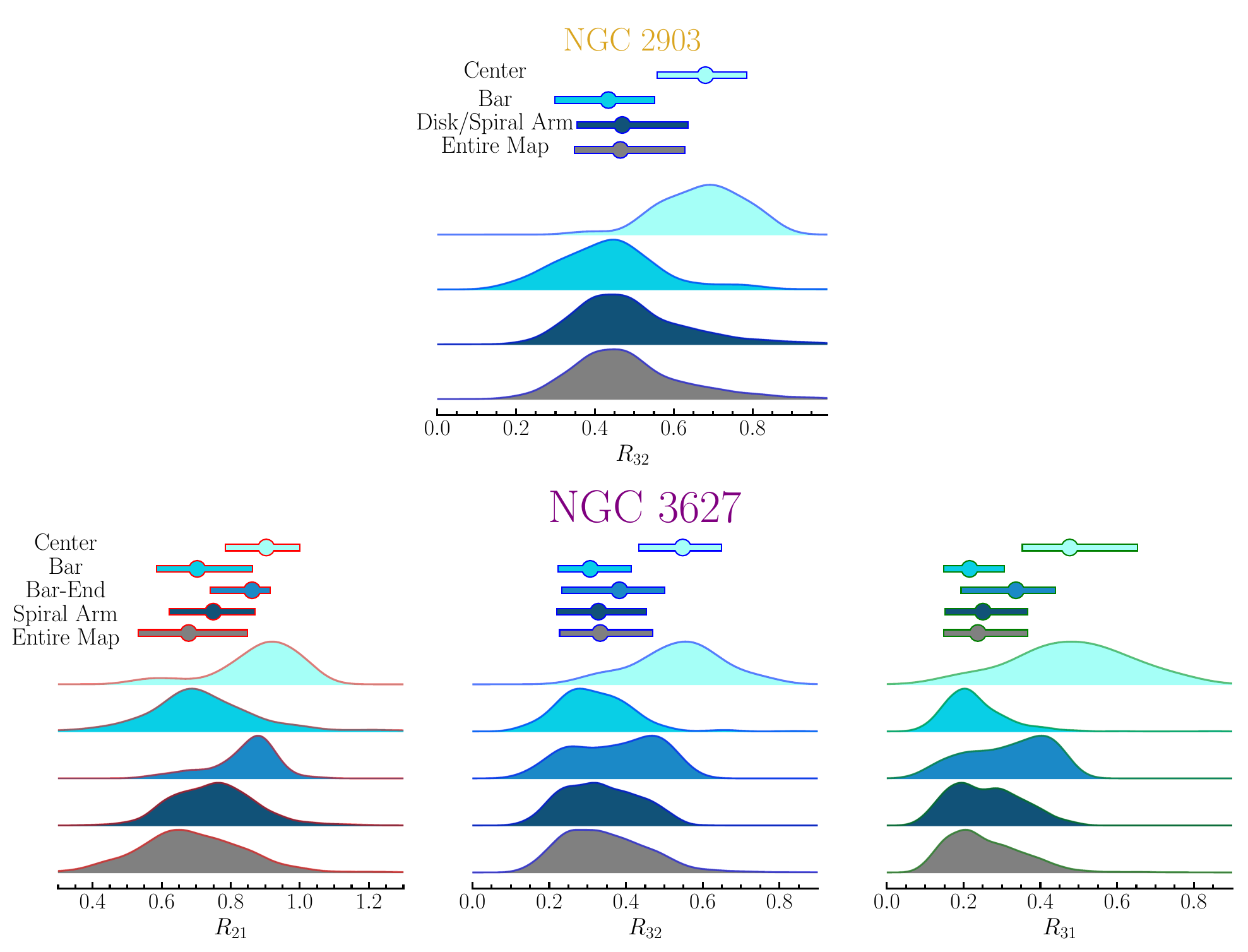}
    \caption{{\bf CO line ratio distribution across different environments in NGC\,2903 and NGC\,3627}  The distribution of $R_{21}$ (\textit{left}), $R_{32}$ (\textit{center}), and $R_{31}$ (\textit{right}) is indicated by a Gaussian kernel distribution function after separating the values of the CO line ratios by different environments. We only include line ratios for which both lines are detected with $\rm S{/}N{>}5$. The median and 16$^{\rm th}$-to-84$^{\rm th}$ percentile range is indicated by the point and bar in the top of the panels (values are listed in \autoref{tab:res_ratio}).}
    \label{fig:ngc3627_ratio_dist}
\end{figure*}
\subsection{CO Line Observations}
We use ALMA to observe low-$J$ transitions of CO.  
Here, we present new CO(3-2) ALMA-ACA data (2019.1.01730.S; PI: J. Puschnig) for two galaxies (NGC\,2903 and NGC\,3627). Both CO(3-2) data sets consist of 7m mosaics at 3\farcs5 angular resolution ($\sim$170\,pc physical scale) and total power observations, with ${\sim}$114\,min total integration time per pointing. At a channel width of 4\,km\,s$^{-1}$, the NGC\,3627 data have an rms of 48\,mK and NGC\,2903 an rms of 32\,mK .
In addition, we use archival CO(1-0) observations for NGC\,3627 (2015.1.01538.S; PI: R. Paladino) and CO(2-1) data from PHANGS-ALMA for both NGC\,3627 and NGC\,2903 \citep{Leroy2021}. 

All three CO transitions are observed as mosaic using a combination of 7-m array, the total power antennas, and (for the \chem{CO}{10} and \chem{CO}{21} lines only) the 12-m array. This ensures that the full spatial information is recovered across the CO disk of NGC 3627. For \chem{CO}{21}, we use the PHANGS-ALMA products directly \citep{Leroy2021}. For \chem{CO}{10} and \chem{CO}{32}, we use the ALMA calibration pipeline to calibrate the interferometric data.  To calibrate the total power, the method described in \citet{Herrera2020} was applied. The imaging and resulting feathering with the total power data were performed using the PHANGS-ALMA pipeline \citep{Leroy2021_pipeline}. 

\begin{figure*}
    \centering
    \includegraphics[width = 0.8\textwidth]{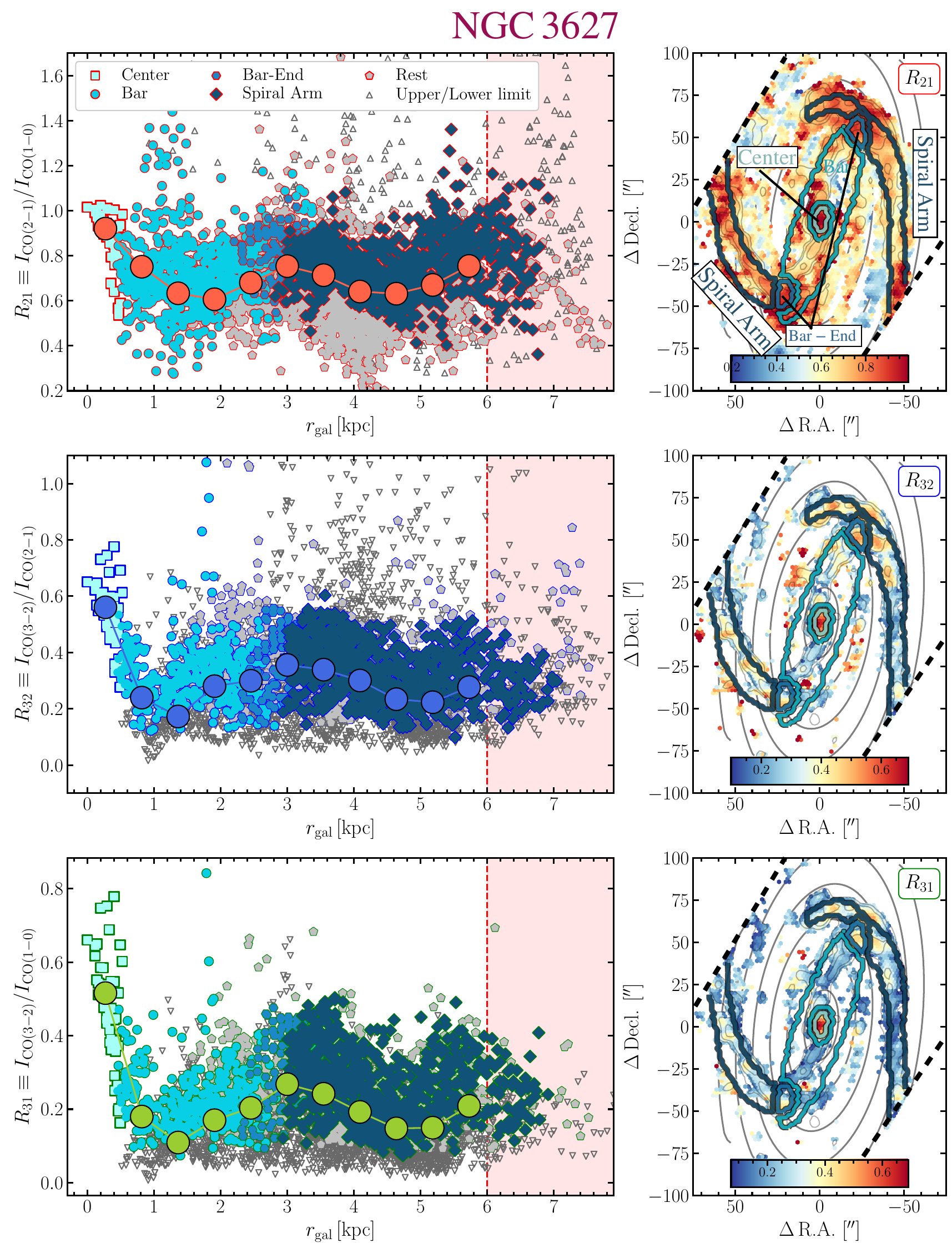}
    \caption{{\bf CO line ratio trends with galactocentric radius in NGC\,3627}. The panels on the left show the radial trend of $R_{21}$ (\textit{top}), $R_{32}$ (\textit{center}), and $R_{31}$ (\textit{bottom}). The red-shaded region shows the radial extent beyond which our sampling is limited by the map size. The large filled circles show the stacked (we stack datapoints irrespective of the different regions).  The sightlines are color-coded by different environmental regions. For reference, the gray ellipses indicate deprojected radial distances of 1, 2, 3, 4, 5, and 6\,kpc. The extent of the different regions is shown in the CO line ratio maps on the right-hand side. For these panels, we only select data points where line intensities have ${\rm S/N}{>5}$ for both the denominator and numerator of the ratio.}
    \label{fig:ngc3627_ratio}
\end{figure*}

\begin{figure*}
    \centering
    \includegraphics[width = 0.8\textwidth]{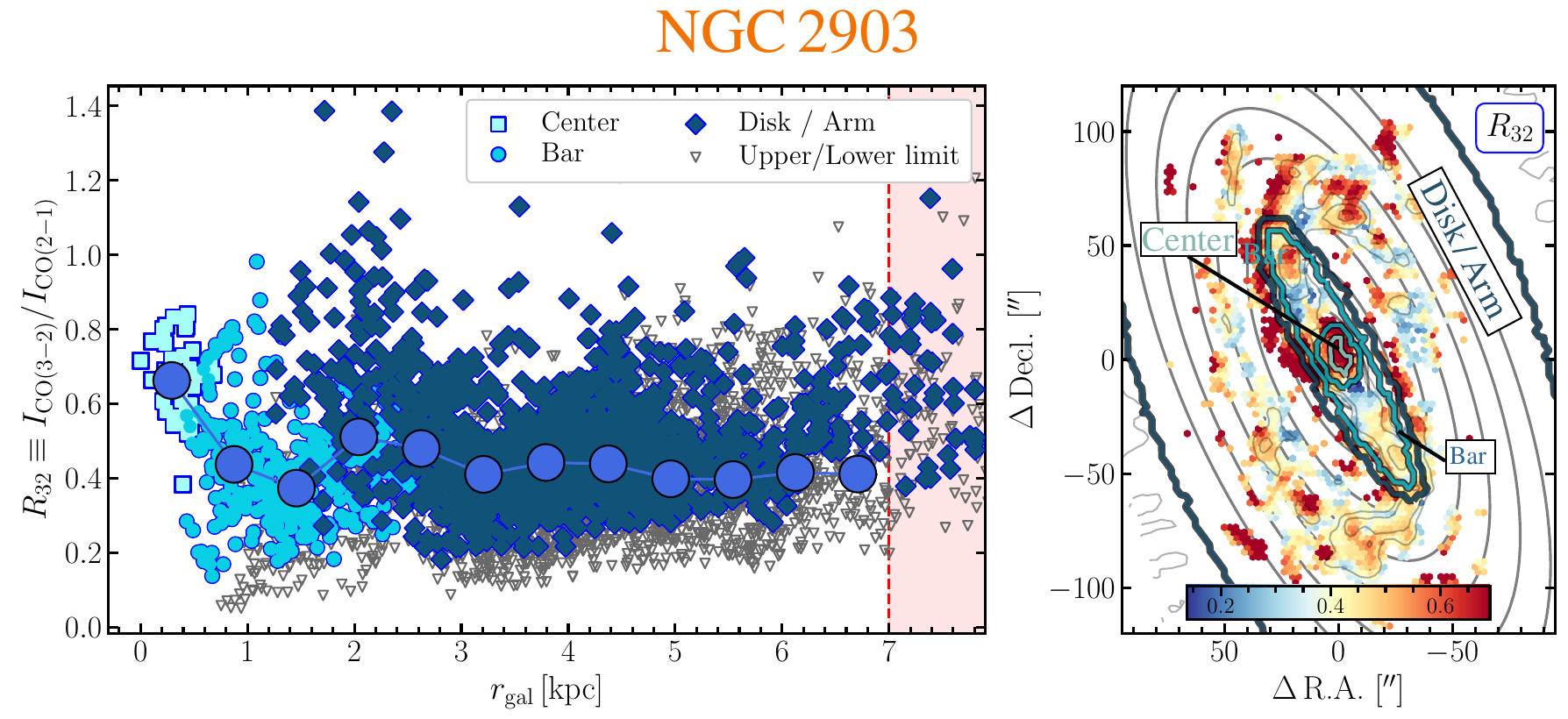}
    \caption{{\bf CO line ratio trends with galactocentric radius in NGC\,2903}. The panels follow the description of \autoref{fig:ngc3627_ratio}.}
    \label{fig:ngc2903_ratio}
\end{figure*}

\begin{figure}
    \centering
    \includegraphics[width=0.8\columnwidth]{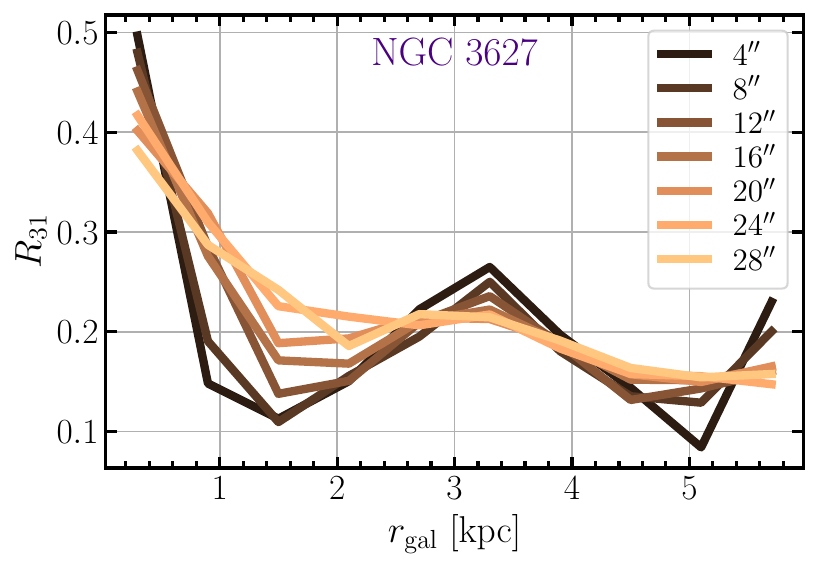}
    \caption{{\bf Radial trend of $R_{31}$ with varying angular resolution.} The stacked line ratio trend is color-coded by the corresponding angular resolution. This panel shows the $R_{31}$ ratio, which shows the strongest radial increase toward the center. }
    \label{fig:varyres}
\end{figure}

\subsection{Data Homogenization and Measurements}
For the subsequent analysis of the cubes, we convolve the observations to a common Gaussian beamsize and regrid to a hexagonal grid following the \texttt{python} pipeline used in \citet{denBrokClaws}. \hyperref[fig:ngc3627_map]{Figure~\ref*{fig:ngc3627_map}} and \autoref{fig:ngc2903_map} show the homogenized data set of the observed CO rotational transition lines for the two galaxies, each regridded, convolved to 4$''$ and only showing the emission within the field of view of the CO(1-0) (NGC\,3627) and CO(3-2) (NGC\,2903) observations. In addition, both figures show an optical HST composite image for reference and present the ancillary data sets used in this analysis. 

The line ratios are computed by integrating each line and calculating the associated uncertainty over an identical velocity window. The velocity window for each line of sight is picked based on the presence of \chem{^{12}CO}{21} emission in the data cube. The mask generally has a width of a few 10s of km\,s$^{-1}$ to 200\,km\,s$^{-1}$ in the center.  We use \chem{^{12}CO}{21} because it is the highest S/N data cube. Then the line ratios are calculated from the ratios of the integrated intensities, and uncertainties on the line ratio are estimated from error propagation from the underlying velocity-window-matched integrated intensities. Among all lines of sight, we identify a set of high signal-to-noise pixels where $\rm S/N > 5$ for both lines of each ratio\footnote{We note that this threshold is higher than the commonly adopted $\rm S/N{>}3$. To minimize any bias by noise, we increased the threshold to five, following \citet{denbrok2021}.}. We will highlight these in plots but also note that particularly for the line ratios involving CO(3-2), the sightlines with S/N${>}5$ contribute around ${\sim}$95\% of the total intensity. We account for this incompleteness in our analysis by using spectral line stacks (which incorporate all sightlines irrespective of their $\rm S/N$ value).  

We perform spectral line stacking following the method described in \citet{Schruba2011, Caldu-Primo2013, Neumann2023}. By stacking, we velocity-normalize and subsequently average individual spectra over all sightlines. Since the noise does not add coherently, this method improves the signal-to-noise (i.e. the noise will decrease), and hence allows us to recover emission from the low signal/noise lines of sight. In the following analysis, any trend lines we fit to the data are linear regression fits to stacked ratios. By performing the trend and correlation analysis on the stacked spectra, we also mitigate the potential bias introduced by our choice of S/N cut for the individual lines of sight.

\subsection{Additional multiwavelength data sets}
In the subsequent analysis, we use additional data sets besides the (sub)mm CO line observations. Here, we briefly describe the origin of the supplementary data sets. 
\subsubsection{JWST mid-IR observations}

NGC 3627 is one of the 19 galaxies that are included in the
PHANGS–JWST Cycle 1 treasury program using the Near
Infrared Camera’s (NIRCam) and MIRI’s
broadband filters \citep[project-ID 02107;][]{Lee2022}. 
We employed the publicly available reduction pipeline and further calibration and reduction steps described in \citet{Lee2022} to obtain science-ready data products (PHANGS-JWST version v0.8).  
For the subsequent analysis, we use 7.7\,$\mu$m and 21\,$\mu$m band observations. The 7.7\,$\mu$m emission traces ionized polycyclic aromatic hydrocarbon (PAH) emission and, in contrast, the 21\,$\mu$m band traces emission from small and hot dust grains \citep{Lee2022}. Therefore, these bands allow us to trace indirectly the heating sources, which we can correlate to the CO excitation via the line ratios. The native angular resolution is ${\sim}$0.3$''$ for the 7.7\,$\mu$m and ${\sim}0.7''$ for the 21\,$\mu$m band. We convolved these to a Gaussian beam of 4$''$ to match the resolution of the CO line transitions using the kernel constructed following \citet{Aniano2011}. At the convolved resolution, the rms is ${\sim}0.4$\,MJy/sr for the 7.7\,$\mu$m and ${\sim}0.1$\,MJy/sr for the 21\,$\mu$m band.
\subsubsection{MUSE H\texorpdfstring{$\alpha$}{Lg}}

NGC\,3627 was observed with the Multi-Unit Spectroscopic Explorer (MUSE)/Very Large Telescope (VLT) instrument as part of the PHANGS-MUSE survey \citep{Emsellem2022}.
The 3D cube has an angular resolution of 1\farcs5. To estimate the SFR surface density, the H$\alpha$ emission is extracted from the MUSE cube. For the final analysis, we smooth the H$\alpha$ map to 4$''$ using a Gaussian kernel.
Furthermore, the H$\alpha$ emission is extinction corrected using the measured Balmer decrement assuming case B recombination. This map is then converted to the SFR surface density following the prescription by \citet{Calzetti2007}. We note that with this description, we do not distinguish between diffuse ionized gas and HII  region emission \citep[see][]{Belfiore2022}.

\subsubsection{Environmental masks}
To study environmental dependencies, we categorize the individual sightlines using the environmental masks (see footprints in \autoref{fig:ngc3627_ratio} and \autoref{fig:ngc2903_ratio}) from \citet{Querejeta2021}. Using \textit{Spitzer} IRAC 3.6\,$\mu$m images from S$^4$G \citep{Sheth2010}, the disks and bulges are identified via 2D photometric decompositions \citep{Salo2015}. The size and orientation of the bar are visually derived. The spiral arms are defined as log-spiral functions fitted to the NIR images (following \citealt{Herrera-Endoqui2015}), with a width assigned based on the PHANGS-ALMA \chem{^{12}CO}{21} intensity distribution. The bar ends are defined based on these masks as the area of overlap between the bar footprint and the innermost end of spiral arms.
We refer to a more detailed description of how these environmental masks are constructed in \citet{Querejeta2021}.

\begin{figure}
    \centering
    \includegraphics[width = 0.9\columnwidth]{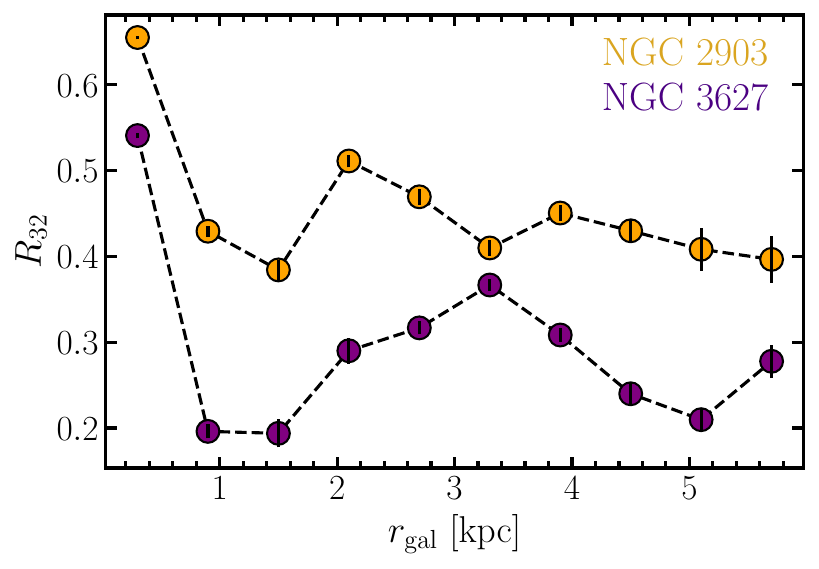}
    \caption{{\bf $R_{32}$ across NGC\,2903 and NGC\,3627}. The panel shows the radial stacked trend of $R_{32}$ for NGC\,3627 (purple) and NGC\,2903 (orange). We find that both galaxies show a similar trend, with a clear increase toward the center. We measure a systematic offset between the trends in the two galaxies of $\Delta R_{32}\approx0.1$--0.2. The bar on each point represents the 1$\sigma$ error. }
    \label{fig:r32_galcomp}
\end{figure}





\section{Results}
\label{sec:result}

Since we have the three lowest-$J$ transition maps for NGC\,3627 and a wealth of ancillary data sets, including MUSE H$\alpha$ and JWST MIR maps, we focus on this galaxy in the first part of the results. We then compare the $R_{32}$ line ratio trends with NGC\,2903. If not otherwise specified, the results refer to line ratio trends and measurements in NGC\,3627. 

\subsection{Variation of the CO Line Ratios across NGC 3627}

{\renewcommand{\arraystretch}{1.5}
\begin{table}
    \centering
    \begin{threeparttable}
    
    \caption{CO ratio summary }
    \label{tab:res_ratio}

    \begin{tabular}{l  c }\hline
         NGC\,2903 & $R_{32}$   \\ \hline\hline
         Center&  $0.68_{-0.11}^{+0.10}$\\
         Bar & $0.43_{-0.10}^{+0.09}$ \\
         Disk / Arms & $0.47_{-0.08}^{+0.13}$ \\ \hline
         Entire Map &$0.46_{-0.08}^{+0.14}$\\\hline
    \end{tabular}
    \vspace{1cm}
    \begin{tabular}{l c c c }\hline
          NGC\,3627 & $R_{21}$ & $R_{32}$ & $R_{31}$  \\ \hline\hline
         Center& $0.90_{-0.11}^{+0.09}$ & $0.54_{-0.10}^{+0.08}$ & $0.48_{-0.12}^{+0.14}$\\
         Bar  &$0.70_{-0.09}^{+0.12}$ & $0.31_{-0.06}^{+0.07}$ &$0.22_{-0.04}^{+0.09}$ \\
         Bar--End &  $0.86_{-0.11}^{+0.04}$ & $0.38_{-0.11}^{+0.06}$ & $0.34_{-0.09}^{+0.07}$\\
         Spiral Arms & $0.75_{-0.11}^{+0.09}$ & $0.33_{-0.08}^{+0.08}$ & $0.25_{-0.08}^{+0.08}$\\ \hline
         Entire Map &$0.67_{-0.11}^{+0.13}$&$0.33_{-0.08}^{+0.09}$&$0.24_{-0.09}^{+0.10}$\\\hline
    \end{tabular}
    
    \begin{tablenotes}
      \small 
      \item {\bf Notes:} The Table lists the median and $16^{\rm th}$ and $84^{\rm th}$ percentile per region including all $\rm S/N{>}5$ pixel-based values. For \chem{^{12}CO}{10}, this includes 98\% of the total flux, for \chem{^{12}CO}{21} 97\%, and for \chem{^{12}CO}{32} 95\%.
    \end{tablenotes}
    \end{threeparttable}
    
\end{table}
}

We show the distribution of values of $R_{21}$, $R_{32}$, and $R_{31}$ across different environments, as defined by the environmental masks \citep{Querejeta2021}, in {{NGC\,2903 ($R_{32}$ only) and}} NGC\,3627  in \autoref{fig:ngc3627_ratio_dist}. The ratio maps and the extent of the individual regions are illustrated in the right panels of \autoref{fig:ngc3627_ratio} and \autoref{fig:ngc2903_ratio}. As shown in \autoref{fig:ngc3627_ratio_dist}, the \emph{Center} {{region of both galaxies has systematically higher CO line ratio values. For this environment, the median is above the 16$^{\rm th}$-to-84$^{\rm th}$ percentile range of values of the other regions for $R_{21}$,  $R_{32}$, and $R_{31}$. For NGC\,3627, also}} the \emph{Bar-End} regions shows, on average, systematically higher $R_{21}$, $R_{32}$, and $R_{31}$ values {{over the galaxy-wide distribution. We note that all three CO line ratios in the \emph{Bar-End} region exhibit a bimodal distribution. This is likely due to uncertainties of the environmental mask, which leads to the inclusion of pixels from the other environments in the \emph{Bar-End} region}}. 

Compared to the galaxy-wide median, $R_{21}$ in the center is $30\%$ larger, $R_{32}$  50\%, and $R_{31}$ 66\%. \autoref{tab:res_ratio} provides the median and 16$^{\rm th}$-to-84$^{\rm th}$ percentile range. 
Such a finding of higher $R_{21}$ and $R_{32}$ values
is expected based on the trend found on kpc-scales in nearby spiral galaxies \citep{Law2018,Israel2020,denbrok2021, Yajima2021, Leroy2022}. For instance, \citet{Law2018} report 
an increase towards the nucleus. They {{measure values of $R_{21}$ ranging}} from $0.2-0.3$ in the spiral arm to $0.5-0.6$ in the nucleus. The increasing line ratio in the center is likely connected to the central region showing molecular gas at higher densities and temperatures \citep{Sun2020,Teng2022,Liu2023,Teng2023}. However, we note that the values found in the center for the different line ratios are still lower than those found in the nearby (ultra-)luminous infrared galaxies, which show even 20-30\% higher values than our average center medians \citep{MontoyaArroyave2023}. Similarly, the bar ends also constitute extreme regions. Compared to the bar and spiral arm, we measure higher average line CO ratio values for $R_{21}$, $R_{32}$, and $R_{31}$ (increase of ${\sim}10{-}20\%$ compared to the galaxy-wide average; see \autoref{tab:res_ratio}). 

To further study the spatial variation across NGC\,3627, we plot the line ratios as a function of galactocentric radius and indicate the different environments by different markers in the left panels of \autoref{fig:ngc3627_ratio}. The large scatter of $R_{21}$, $R_{32}$, and $R_{31}$ for the central environment\footnote{We differentiate between the center of the galaxy (${<}0.5$kpc) and the central environment, which spans out to the bar ends (${<}2$\,kpc)} is due to a significant decreasing trend of the respective line ratios with galactocentric radius. This ``hook" towards small radii is most prominently seen for $R_{31}$. Looking quantitatively at the line ratios from $R_{\rm gal}=0-1.5\,$kpc, $R_{21}$ decreases radially by ${\sim}35$\% from $0.99\pm0.01$ to $0.65\pm0.02$. The radial decrease in $R_{32}$ within the most central 1.5\,kpc is even stronger with ${\sim}50$\% from $0.62\pm0.01$ to $0.32\pm0.03$. And for $R_{31}$ the decrease amounts to ${\sim}70$\% from $0.65\pm0.01$ to $0.21\pm0.03$.
The fact that the ratios including \chem{^{12}CO}{32} emission show the steepest radial gradient toward the center suggests that the excitation conditions affect this higher-$J$ transition more significantly than the \trans{21} and \trans{10} transitions.


We note that beyond $r_{\rm gal}{>}6\,\rm kpc$, we have incomplete sampling (see red shaded space in \autoref{fig:ngc3627_ratio}; at these radii, sightlines fall outside of our map coverage) and hence we cannot trace the trend in $R_{21}$, $R_{32}$, and $R_{31}$ there. The slight upturn we detect toward higher radii at ${>}5\,$kpc in the stacked trend likely reflects noise and selection bias, as in the outskirts the CO brightness becomes fainter and our radial sampling will be more incomplete. 

Generally, only the central ($r_{\rm gal}{<}2\,\rm kpc$) region (galactic center and bar) shows a clear internal radial variation of the line ratios, with the highest values in the very center of the galaxy. 
In comparison to previous studies, we resolve  enhanced ratios in distinct environments -- the center and bar ends -- that lower resolution would smooth into a continuous declining radial profile \citep[e.g.,][]{denbrok2021}. This is evident in \autoref{fig:varyres}, where the radial trend lines show the $R_{31}$ variation for different angular resolutions, spanning from $4''$ to $28''$. In Appendix~\ref{sec:resol_analysis}, we discuss in more detail the overall line ratio distribution for different angular resolutions in NGC\,3627. Care needs to be taken in particular since the sensitivity will change as a function of angular resolution.

\begin{figure*}
    \centering
    \includegraphics[width = 0.95\textwidth]{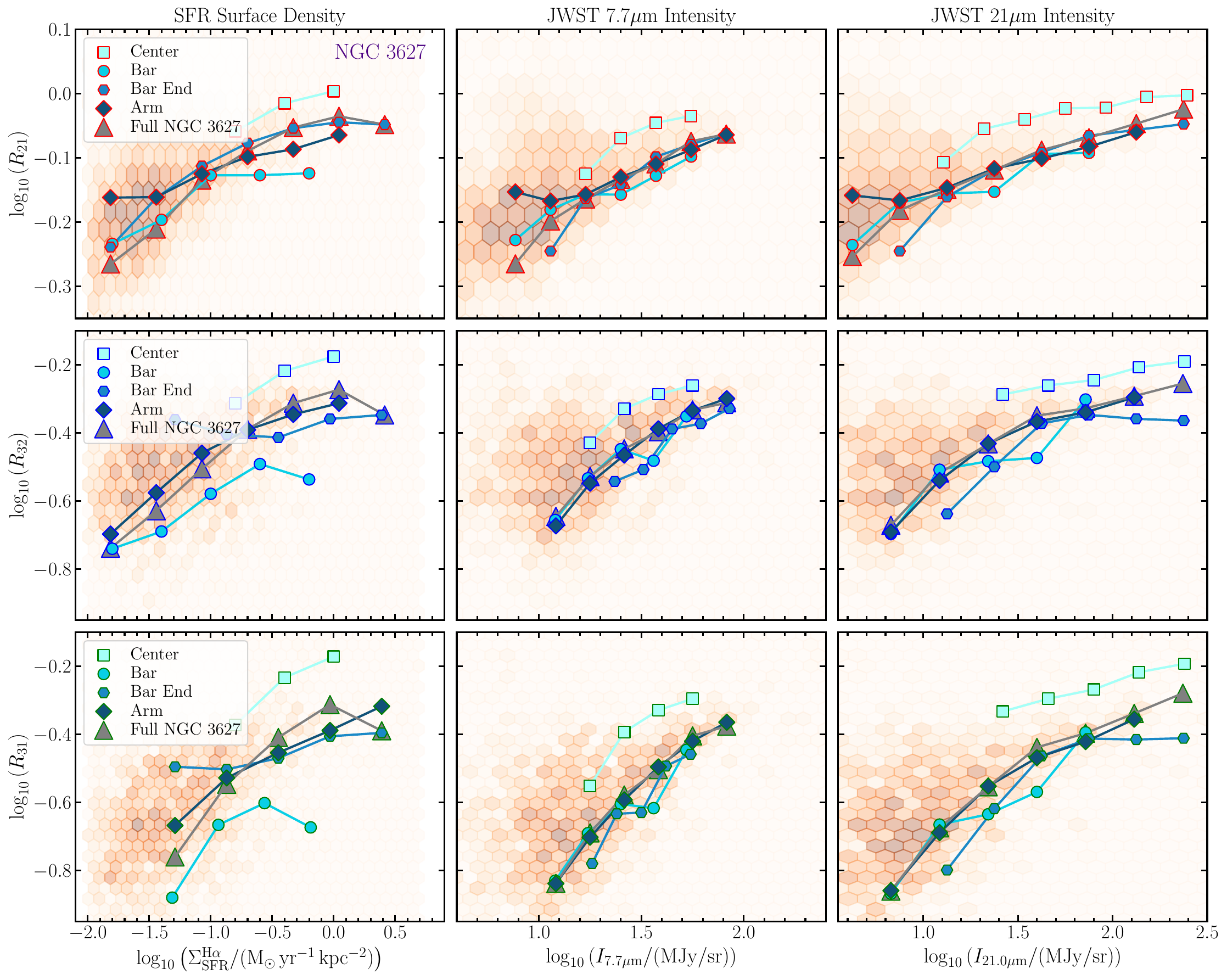}
    \caption{{\bf CO line ratio trends with $\Sigma_{\rm SFR}$ and dust tracers $I_{7.7\,\mu\rm m}$ (PAH) and  $I_{7.7\,\mu\rm m}$ (hot dust)}. The different panels show the trends of $R_{21}$ (\textit{top}), $R_{32}$ (\textit{middle}), and $R_{31}$ (\textit{bottom}) for the different environments. The orange hexagonal bins represent the 2D histogram of $\rm S/N{\ge}5$ sightlines. We note that the stacked trends also include the sightlines $\rm S/N{<}5$.
    } 
    \label{fig:ratio_comp}
\end{figure*}


\begin{figure}
    \centering
    \includegraphics[width = 0.75\columnwidth]{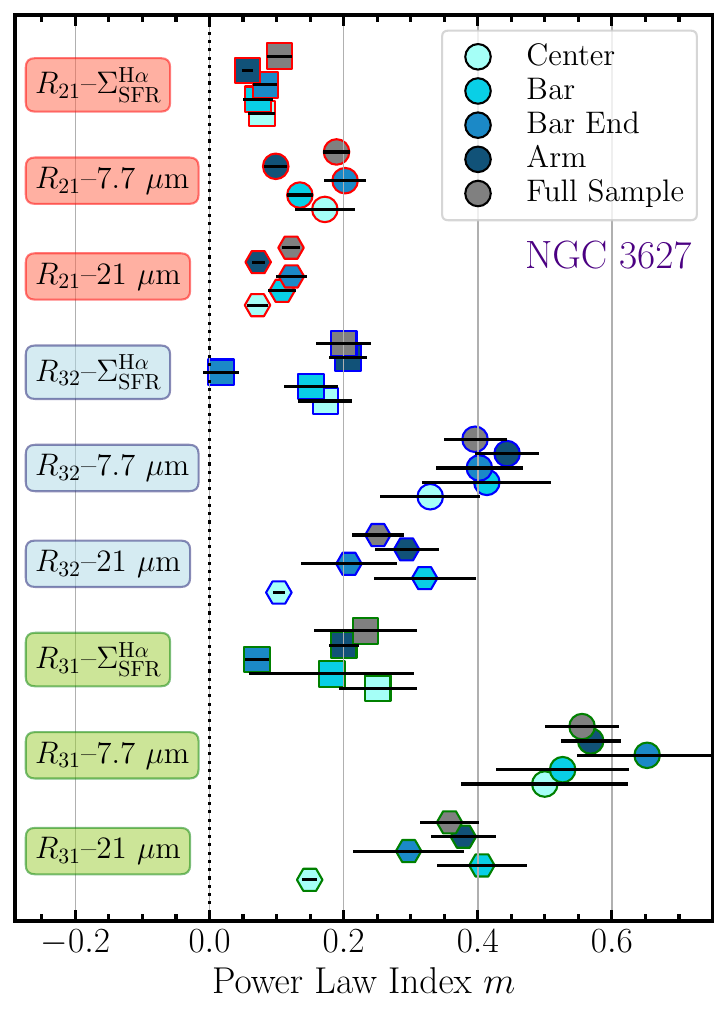}
    \caption{{\bf Index of power-law fits of the CO line ratios.} This figure reports the linear regression coefficient of the stacked trends shown in \autoref{fig:ratio_comp} in log-log space. The points coloring scheme follows the classification by environment in \autoref{fig:ratio_comp}. The steepest trends are generally found with $R_{32}$. While for the correlation with $\Sigma_{\rm SFR}$, the trend is steepest for the center, for the mid-IR bands, we find that the center has a shallower slope respectively. The black line indicates the standard deviation of the linear regression fit.}
    \label{fig:corr_check}
\end{figure}

\begin{figure}
    \centering
    \includegraphics[width = \columnwidth]{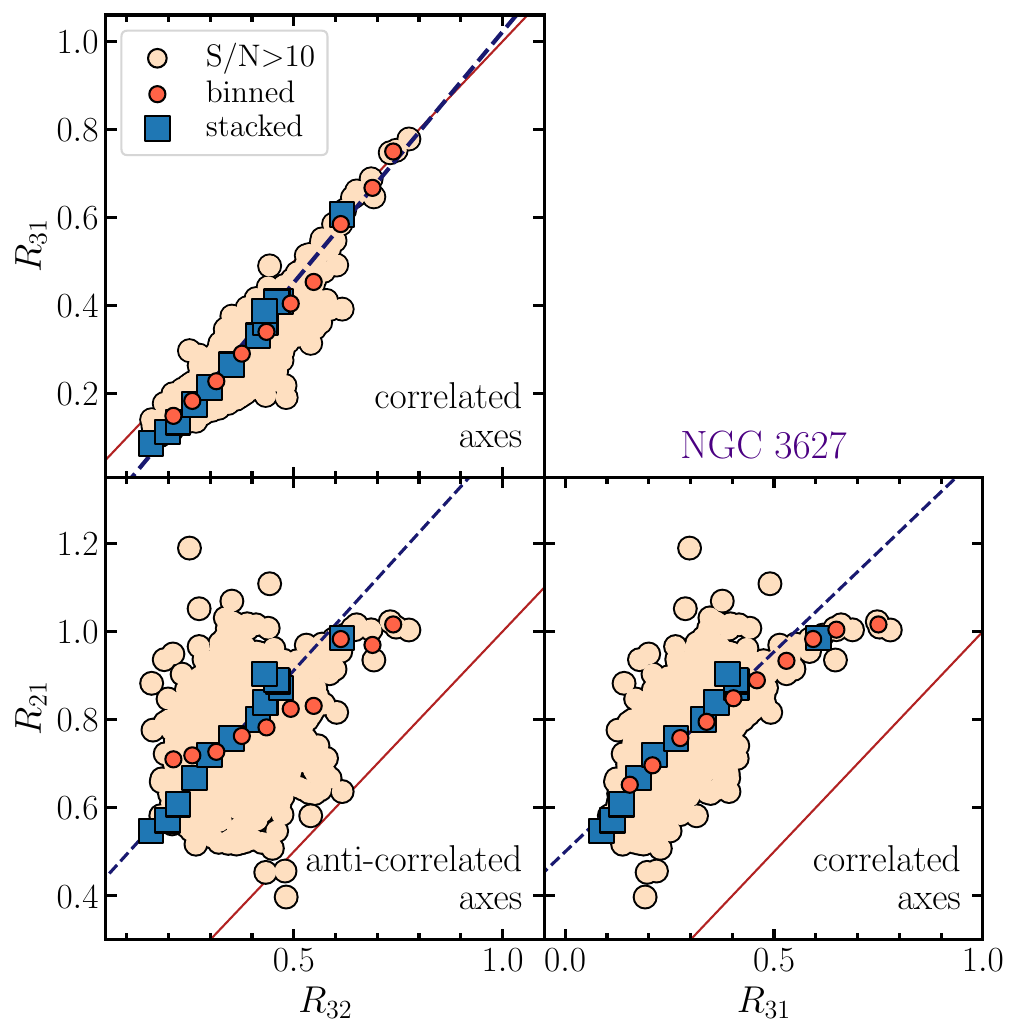}
    \caption{{\bf CO ratio-to-ratio comparison in NGC\,3627}. The panels depict the three permutations of comparing $R_{21}$, $R_{32}$, and $R_{31}$ to each other. The light-colored points have S/N${>}5$ in all lines. The filled circles are the binned trend. The filled squares represent the result when computing the line ratio from the spectral stacks by SFR (using all lines of sight irrespective of S/N). For reference, the red line indicates the 1:1 relation. We perform a linear regression to the stacked data points, represented as a black dashed line. We note that the axes of the different panels are correlated. The sense of the implicit correlation between the plotted variables is indicated in each panel. } 
    \label{fig:ratio_vs_ratio}
\end{figure}



\subsection{\texorpdfstring{$R_{32}$}{Lg} in NGC\,2903}

We measure that overall the $R_{32}$ line ratios are systematically higher in NGC\,2903 than in NGC\,3627 by $\Delta R_{32}\approx0.1-0.2$  across the galaxy's center and disk. More strikingly, looking at the radial trend in both galaxies (\autoref{fig:r32_galcomp}), we detect a very similar pattern: a steep increase of the line ratio toward the center for $r_{\rm gal}{<}1\,$kpc. The disks have flatter trends, with bumps at the radii corresponding to the bar ends.  The central value rises by $\Delta R_{32}\approx 0.3$ from 1 to 0\,kpc radial distance in both sources. Since we only have two rotational-$J$ transitions for NGC\,2903, we cannot perform the same line analysis as for NGC\,3627. The fact that both centers show an increase of the CO line ratio of similar order likely reflects a similar mechanism that enhances higher-J CO emission in such environments (see  discussion in \autoref{sec:modelling}). This comparison highlights the fact that the galaxy centers and bar ends are environments that significantly affect the shape of the CO spectral line energy distribution (SLED). Looking at larger sample sizes, kiloparsec-scale studies report that not all galaxies show such radial increases toward the center \citep[e.g. $R_{21}$,][]{denbrok2021}. 


\begin{table*}
    \centering
    \begin{threeparttable}
    
    \caption{Index and constant offset of power-law fits in NGC\,3627}
    \label{tab:powerLawfit}
    {\small
    
\begin{tabular}{cl|l|llll}
\multicolumn{2}{l|}{}                          & \multicolumn{1}{c|}{Full Sample}   & \multicolumn{1}{c}{Center}          & \multicolumn{1}{c}{Bar}            & \multicolumn{1}{c}{Bar End}        & \multicolumn{1}{c}{Arm}                \\ \hline
\multirow{3}{*}{$R_{21}$} & $\Sigma_{\rm SFR}$ & $0.10{\pm}0.02$ / $-0.05{\pm}0.02$ & $0.08{\pm}0.01$ / $0.008{\pm}0.009$ & $0.07{\pm}0.02$ / $-0.09{\pm}0.02$ & $0.08{\pm}0.02$ / $-0.05{\pm}0.02$ & $0.056{\pm}0.005$ / $-0.066{\pm}0.005$ \\
                          & 7.7 $\mu$m         & $0.18{\pm}0.02$ / $-0.41{\pm}0.02$ & $0.17{\pm}0.04$ / $-0.32{\pm}0.06$  & $0.13{\pm}0.02$ / $-0.34{\pm}0.02$ & $0.20{\pm}0.03$ / $-0.43{\pm}0.04$ & $0.10{\pm}0.09$ / $-0.26{\pm}0.02$     \\
                          & 21 $\mu$m          & $0.12{\pm}0.01$ / $-0.30{\pm}0.02$ & $0.07{\pm}0.01$ / $-0.16{\pm}0.02$  & $0.11{\pm}0.02$ / $-0.28{\pm}0.02$ & $0.12{\pm}0.02$ / $-0.31{\pm}0.04$ & $0.073{\pm}0.007$ / $-0.22{\pm}0.01$   \\ \hline
\multirow{3}{*}{$R_{32}$} & $\Sigma_{\rm SFR}$ & $0.20{\pm}0.04$ / $-0.32{\pm}0.04$ & $0.17{\pm}0.04$ / $-0.16{\pm}0.02$  & $0.15{\pm}0.04$ / $-0.45{\pm}0.04$ & $0.02{\pm}0.02$ / $-0.37{\pm}0.02$ & $0.21{\pm}0.03$ / $-0.28{\pm}0.03$     \\
                          & 7.7 $\mu$m         & $0.40{\pm}0.04$ / $-1.04{\pm}0.07$ & $0.33{\pm}0.07$ / $-0.8{\pm}0.1$    & $0.41{\pm}0.09$ / $-1.1{\pm}0.1$   & $0.40{\pm}0.06$ / $-1.1{\pm}0.1$   & $0.44{\pm}0.04$ / $-1.11{\pm}0.07$     \\
                          & 21 $\mu$m          & $0.25{\pm}0.04$ / $-0.81{\pm}0.06$ & $0.10{\pm}0.01$ / $-0.43{\pm}0.01$  & $0.32{\pm}0.07$ / $-0.9{\pm}0.1$   & $0.21{\pm}0.04$ / $-0.8{\pm}0.1$   & $0.30{\pm}0.05$ / $-0.88{\pm}0.07$     \\ \hline
\multirow{3}{*}{$R_{31}$} & $\Sigma_{\rm SFR}$ & $0.23{\pm}0.07$ / $-0.37{\pm}0.06$ & $0.25{\pm}0.06$ / $-0.16{\pm}0.03$  & $0.2{\pm}0.1$ / $-0.5{\pm}0.1$     & $0.07{\pm}0.02$ / $-0.42{\pm}0.01$ & $0.20{\pm}0.01$ / $-0.38{\pm}0.01$     \\
                          & 7.7 $\mu$m         & $0.56{\pm}0.05$ / $-1.40{\pm}0.08$ & $0.5{\pm}0.1$ / $-1.1{\pm}0.2$      & $0.5{\pm}0.1$ / $-1.4{\pm}0.1$     & $0.7{\pm}0.1$ / $-1.6{\pm}0.2$     & $0.57{\pm}0.04$ / $-1.4{\pm}0.1$       \\
                          & 21 $\mu$m          & $0.36{\pm}0.04$ / $-1.09{\pm}0.07$ & $0.15{\pm}0.08$ / $-0.54{\pm}0.01$  & $0.41{\pm}0.06$ / $-1.17{\pm}0.09$ & $0.30{\pm}0.08$ / $-1.0{\pm}0.1$   & $0.39{\pm}0.05$ / $-1.11{\pm}0.07$    
\end{tabular}

    }
    \begin{tablenotes}
      \small 
      \item {\bf Notes:} The values indicate $m$/$q$, where $m$ is the power law index and $q$ the constant offset. See \autoref{fig:ratio_comp} for the correlation plotted against each other. The coefficients are derived assuming a power-law correlation for the stacked line ratio trends. The standard error of the slope and intercept is derived under the assumption of residual normality.
    \end{tablenotes}
    \end{threeparttable}
    
\end{table*}

\begin{table}
    \centering
    \begin{threeparttable}
    
    \caption{Ratio-to-ratio comparison for NGC\,3627}
    \label{tab:ratio_vs_ratio}
    
    \begin{tabular}{l l cc  c c }\hline
          $x$ & $y$ & $m$  & $q$&$p$  & scatter [dex] \\ \hline\hline
         $R_{32}$ & $R_{31}$& $1.1$ & $-0.12$ & $1\times 10^{-12}$& $0.062$\\
         $R_{32}$& $R_{21}$& $1.0$ & $0.39$& $4\times 10^{-9}$&$0.15$ \\
         $R_{31}$& $R_{21}$& $0.91$  & $0.50$ &$1\times 10^{-10}$&$0.072$\\\hline
    \end{tabular}
    
    \begin{tablenotes}
      \small 
      \item {\bf Notes:} See \autoref{fig:ratio_vs_ratio} for the ratios plotted against each other. The coefficients are derived assuming a linear correlation $y=mx+q$ of the stacked ratio-to-ratio trend. The $p$-value describes the probability assuming no correlation (null hypothesis). The scatter shows the standard deviation of the $\rm S/N>5$ data points with respect to the best-fit line.
    \end{tablenotes}
    \end{threeparttable}
    
\end{table}

{\subsection{Tracing varying CO excitation conditions}}
While CO excitation and the resulting CO line ratio observations likely depend on parameters such as the gas density, temperature, and optical depth distributions, all of which are difficult to observe directly, studies have found a close correlation between these quantities and the star formation rate (SFR) surface density \citep[e.g., ][]{Schmidt1959,Kennicutt1989,Narayanan2012, Jiang2019}. On kpc-scales,  the SFR surface density, $\Sigma_{\rm SFR}$, correlates with the CO line ratio $R_{21}$ \citep[][]{Bayet2009, Koda2012, Egusa2022, Leroy2022, Leroy2023_JWST}. 
To test the correlation within the different environments in NGC\,3627, we plot the $R_{21}$, $R_{32}$, and $R_{31}$ line ratio as a function of the SFR surface density and the dust tracers in \autoref{fig:ratio_comp} (panels in left column). Within NGC\,3627, we cover a dynamical range of about 2\,dex of SFR surface density at high angular resolution (4''). Compared to the other line ratios, for $R_{21}$, we find a consistently lower power law slope of $m\le0.1$ for the different regions and $m=0.10$ for the full sample. 
For $R_{32}$, the slope $m$, tends to be larger than $R_{21}$ with $m=0.15-0.21$, with the exception of the \emph{Bar-End} region, where the slope is lower. The \emph{Bar-End} region shows a near-flat trend. However, our dynamical range in SFR is limited. The slope for the full sample is $m=0.20$.  We also note that, while the slope is similar, the central environment has an offset of the trend line with respect to the full sample and other environments by $\le0.1\,$dex, indicating a change in molecular gas conditions in the center compared to the rest of the galaxy. Generally, we find a clear correlation between SFR surface density and the line ratio across all the different environments, hinting at an underlying link between SFR surface density, the physical gas conditions (temperature and density), and the resulting observed CO line ratios in accordance with previous studies \citep[e.g.,][]{Bayet2009, Lamperti2020} .  


In \autoref{fig:ratio_comp}, we also contrast the CO line ratios to the mid-IR 7.7\,$\mu$m (panels in middle column) and 21\,$\mu$m (panels in right column) emission. Both mid-IR bands showed a correlation with the line ratios on kpc scale \citep{Leroy2023_JWST}. The  7.7\,$\mu$m band traces ionized PAH and the 21\,$\mu$m band emission from hot small dust grains \citep[e.g.,][]{Chastenet2023,Leroy2023_JWST}. We find an increasing trend for all three CO line ratios with the two mid-IR emission bands. Each panel presents stacked line ratio trends separated by the different environmental regions. We find a steeper correlation slope between the line ratio and the 7.7$\,\mu$m band than the 21$\,\mu$m band emission by a factor of ${\sim}2$. While the bar-end region shows a flat trend when compared to the SFR surface density for $R_{32}$ and $R_{31}$, the correlation with the mid-IR does not show any flat trend. Again, the center environment has a trend that is offset toward higher line ratios by $\le0.1$\,dex. In contrast, the ratios show a similar dependence on the SFR surface density, implying that change in SFR similarly impacts higher and lower-$J$ transitions. 

\hyperref[fig:corr_check]{Figure~\ref*{fig:corr_check}} provides a compilation of the indices of the power-law correlation fit for the CO line ratios in NGC\,3627 with $\Sigma_{\rm SFR}$ and the 7.7\,$\mu$m and 21.0\,$\mu$m bands. The figure indicates that all three line ratios show the steepest relation with the  7.7\,$\mu$m IR band emission. This is possibly because the 7.7\,$\mu$m emission probes both higher density and the heating rate, while the 21\,$\mu$m traces mostly the heating rate only. Compared to the full sample fit, we note that for all three quantities the center is significantly brighter at a fixed ratio than the overall average trend. The power-law indices for the different correlations with the CO line ratios are listed in \autoref{tab:powerLawfit}.

{\subsection{Ratio to ratio comparison}}
The three CO lines that we study depend differently on density and temperature. { For instance, as we will discuss in more detail in \autoref{sec:modelling}, we find that $R_{21}$ is more sensitive to changes in temperature, particularly at $T_{\rm kin}{>}30\,\rm K$.} It has been relatively rare to measure all three lines at high resolution for a whole target, and comparing the ratios is therefore of interest. \hyperref[fig:ratio_vs_ratio]{Figure~\ref*{fig:ratio_vs_ratio}} illustrates the three permutations of contrasting $R_{21}$, $R_{32}$, and $R_{31}$ to each other. Since such a comparison is sensitive to noise effects, we stack the data to show trends more effectively. We also report the trend based on the spectral line stacks using the SFR surface density (shown in blue) derived from the MUSE H$\alpha$ observations.  All three panels depict a positive linear correlation. We note that in the case of the $R_{31}$-to-$R_{32}$ (top panel) and $R_{31}$-to-$R_{21}$ (bottom right panel) comparison, we are dealing with correlated axes making it  difficult to filter out the intrinsic ratio trend. However, in the case of the $R_{32}$-to-$R_{21}$ (bottom left panel) comparison, the axes are actually anti-correlated. Nonetheless, we detect a positive correlation between the two line ratios. The stacked trend follows the binned trend. In the bottom two panels, the trends with $R_{21}$ on the $y$-axis show a flattening at $R_{21}{\sim}1$, which corresponds to data points at the center of the galaxy, where $I_{\chem{^{12}CO}{32}}$ increases most significantly relative to the other two lines.{ Generally, the observed ratio-to-ratio correlations suggest that the $J{=}3{\rightarrow}2$ transition is more sensitive to changes in the excitation conditions, which is likely linked to its higher critical density with respect to the lower $J$ transitions. Extracting the particular driver for the excitation condition (higher temperature, change in opacity or density) is complex and requires contrasting observed and simulated line ratios, based on different molecular gas conditions.  We provide a more detailed discussion of the possible explanations for the ratio-to-ratio correlations, as well as the scatter, in \autoref{disc:overall}}.  \autoref{tab:ratio_vs_ratio} lists the linear regression fit coefficients for the stacked line ratio-to-ratio trend. 
\medskip


\section{Discussion}
\label{sec:discussion}
\begin{figure}
    \centering
    \includegraphics[width = 0.85\columnwidth]{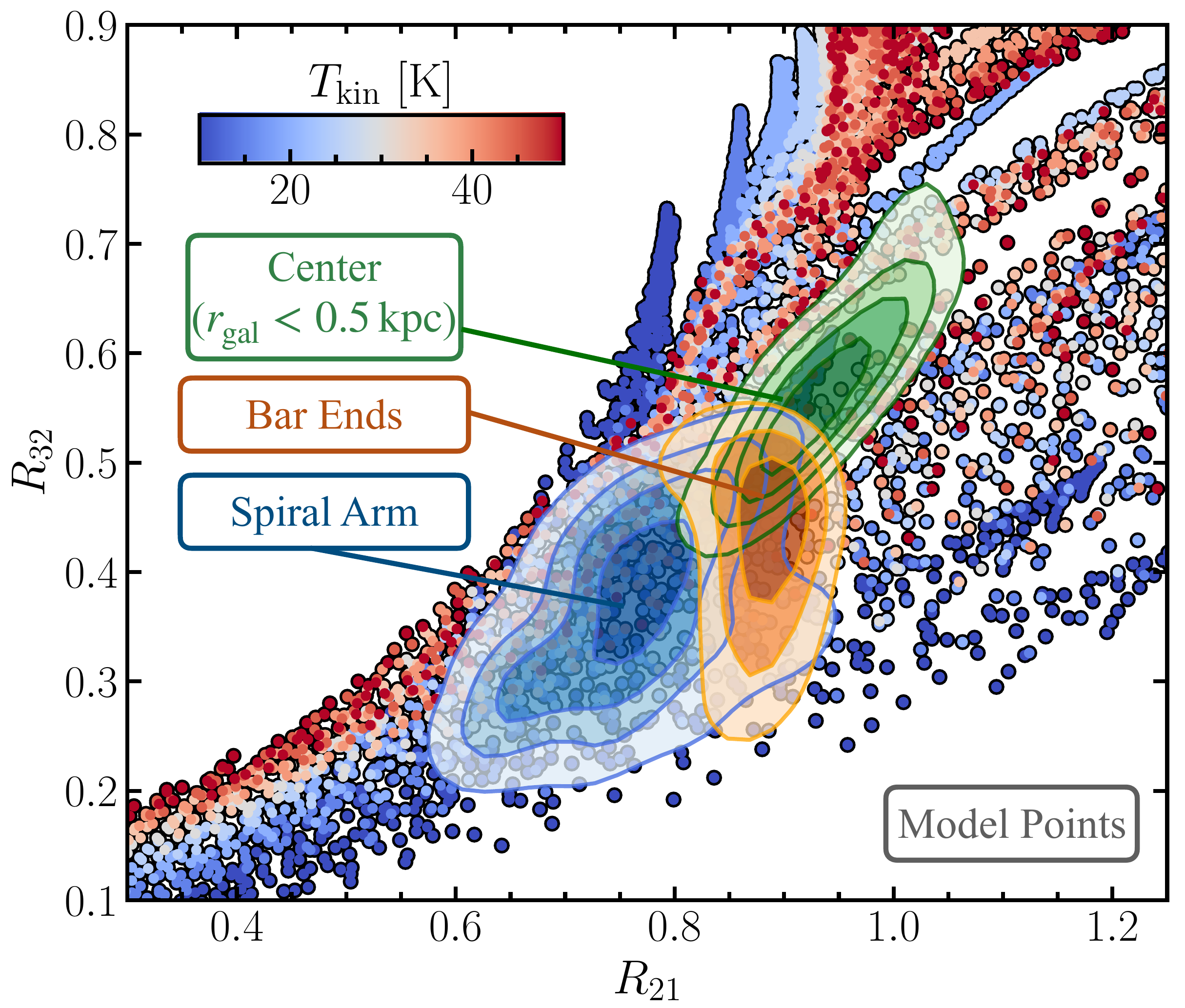}
    \caption{{\bf Comparison of measured and model ratios.} The points color-coded by their kinetic temperature illustrate the CO line ratios produced by model calculations from \citet{Leroy2022}. The contours show the 80\% inclusion distribution of the CO line ratios, measured in NGC\,3627 , of the center region ($r_{\rm gal}{<}0.5$\,kpc) in orange, the spiral arm in green, and bar end region in blue.}
    \label{fig:model_comp_ratios}
\end{figure}

\begin{figure*}
    \centering
    \includegraphics[width = 0.8\textwidth]{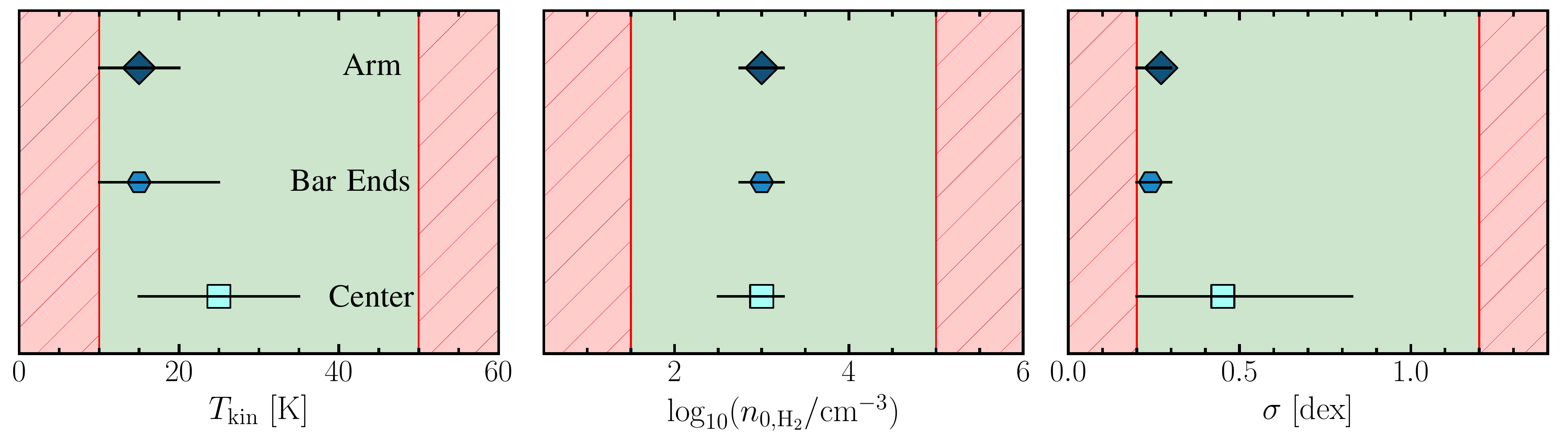}
    \caption{{\bf Distribution of temperature, mean H$_2$ volume density, and density dispersion} We compute for each pixel the temperature, $T_{\rm kin}$, mean volume density, $n_{\rm H_2}$, and density dispersion, $\sigma$ using the maximum likelihood approach. The markers indicate the median value for spiral arm, bar ends, and center. The 16$^{\rm th}$-84$^{\rm th}$ percentile range is characterized by the black line. The ranges of values covered by the model grid are highlighted in green. The value range outside the model grid is shaded in red.}
    \label{fig:model_results}
\end{figure*}

\begin{figure}
    \centering
    \includegraphics[width = 0.85\columnwidth]{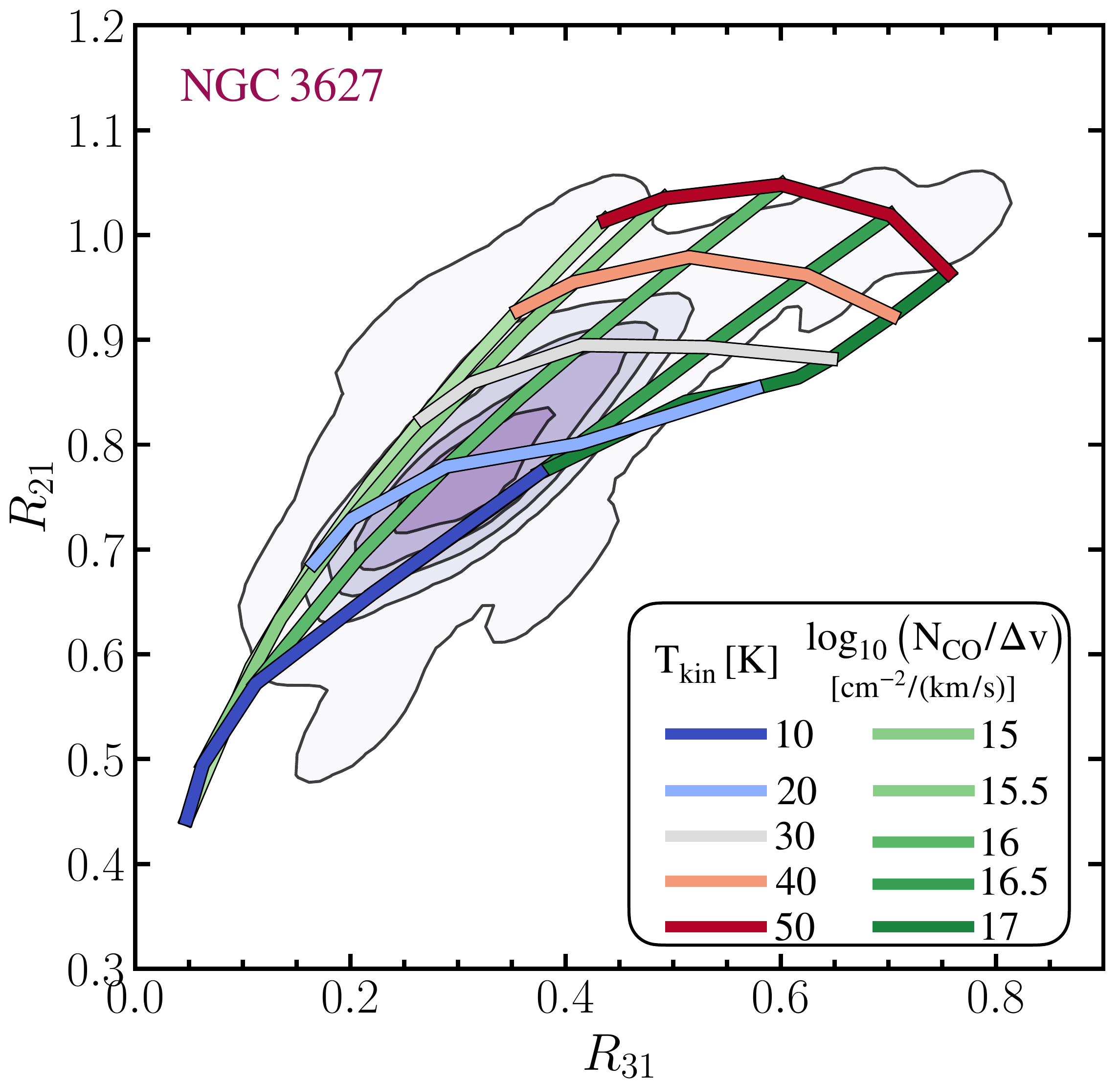}
    \caption{{\bf Model grid and CO line ratio distribution.} Based on the model results in \autoref{sec:modelling}, we plot here the modelled $R_{31}$ versus $R_{21}$ assuming a constant mean volume density of $n_{0,\rm H_2}=10^3\,$cm$^{-3}$ and $\sigma=0.3$. The increase in temperature (blue to red) describes well the linear correlation of the line ratios observed in NGC\,3627 (indicated by the purple contours, which indicate 20\%, 40\%, 60\%, 80\%, and 100\% inclusion). The spread is characterized by changes in $N_{\rm CO}/\Delta v$, which reflects variation in optical depth.}
    \label{fig:model_grid}
\end{figure}

\subsection{Modelling CO line ratios}
\label{sec:modelling}
Past studies have reported on a theoretical dependence of the observed line ratios $R_{21}$, $R_{32}$ on key physical properties of the molecular gas \citep[e.g.,][]{Bolatto2013, Shirley2015, Leroy2017, Penaloza2018,Leroy2022}. These input parameters to the model include the kinetic temperature, $T_{\rm kin}$, the mean density of H$_{2}$, $n_{\rm 0, H_{2}}$, the width (in dex) of the H$_{2}$ density dispersion, $\sigma$, and the CO column density per line width $N_{\rm CO}/\Delta v$. In the subsequent analysis, we employ the models calculated by \citet{Leroy2022}. The particular advantage of these models is that we mimic a more realistic mixture of densities based on combining RADEX models, instead of just a single or two-zone molecular ISM. A density dispersion naturally follows from a turbulent ISM description in nearby galaxies. Following such turbulent models of star formation, the density is described as a log-normal distribution \citep[e.g.][]{Federrath2012, Padoan2014}. In addition, the model includes a power-law tail component at high densities that is produced by gravitational collapse \citep[e.g.,][]{Girichidis2014}. 
 The predicted  \chem{^{12}CO}{10}, \trans{21}, and \trans{32} emission are computed using \texttt{RADEX} \citep{vanTak2007} and data from the Leiden Atomic and Molecular Database \citep[LAMDA; ][]{Schoier2005}. While these models assume a  density variation, the $N_{\rm CO}/\Delta v$ remains fixed across all zones per model point (but the exact value can vary for each data grid point across the galaxy). We adopt a  $\Delta v$ based on the CO(2-1) velocity width from our ALMA observations. We estimate $N_{\rm CO}$ using the measured CO(2-1) intensity and a metallicity dependent $\alpha_{\rm CO}$ from \citet{Sun2022}. Dividing the CO intensity by the conversion factor yields a measure of  $N_{\rm H_2}$. Therefore, to convert to $N_{\rm CO}$, we assume a CO abundance of $[\rm CO/H_2]=10^{-4}$. For more details on the general model description and generation, we refer the reader to Appendix A of \citet{Leroy2022}.



At various locations across NGC\,3627, we measure above the galaxy-wide average CO line ratios, particularly near the center. Here, we assess what physical conditions can reproduce such behavior in the observed line ratios. We specifically investigate sightlines from the center of the galaxy ($r_{\rm gal}{\le}0.5\,$kpc) and the arm regions (\emph{Arm (North+South)}), excluding the bar ends. In \autoref{fig:model_comp_ratios}, we compare the measured line ratio distribution of the two distinct environments to the model calculations. To assess the distribution of physical parameters that best describe the observed CO line ratio distribution, we perform a $\chi^2$ analysis on the underlying lines of sight of the different regions. We combine the resulting (area-normalized) probability density functions (PDFs) to obtain a  PDF per environment and compute $\chi^2$ as 
\begin{equation}
    \chi^2(\theta) = \sum_{i=1}^n \frac{\left(S_i^{\rm mod}(\theta)-S_i^{\rm obs}\right)^2}{\rm unc_i^2},
\end{equation}
with $\theta=(n_{0,\rm H_2}, \sigma, N_{\rm H}/{\rm d}v, T_{\rm ex})$, $S_i$ for the (modelled or observed) $R_{21}$ and $R_{32}$ line ratio, $n$ the number of transition lines and $\rm unc_i$ the measured uncertainty.
The subsequent PDF is computed following:
\begin{equation}
    P(S_{\rm model},\theta) = \frac{1}{\prod_i\left(\sqrt{2\pi\cdot\rm unc_i}\right)}\cdot e^{\left(-\frac{1}{2}\chi^2(\theta)\right)}
\end{equation}

    
    
    
    

For each line of sight, we determine the temperature, density, and lognormal density dispersion using the maximum likelihood. Appendix \ref{app_model} shows the $\chi^2$ minimization result for a single line of sight in the center and spiral arm. In \autoref{fig:model_results}, we present the median (and 16$^{\rm th}$-84$^{\rm th}$ percentile range) temperature, volume density, and density dispersion across all sightlines for three distinct environments: the center, bar ends, and the spiral arm region. We find that the center region is warmer ($T_{\rm kin}=25^{+10}_{-10}\,\rm K$)  than the bar ends ($T_{\rm kin}=15^{+10}_{-5}\,\rm K$) and spiral arm ($T_{\rm kin}=15^{+5}_{-5}\,\rm K$). In contrast, the derived mean volume densities, as traced by the $^{12}$CO, do not significantly differ between the environments, where we find that the center has $\log_{10}(n_{\rm 0,H_2})=3^{+0.25}_{-0.50}$, the bar ends have $\log_{10}(n_{\rm 0,H_2})=3^{+0.25}_{-0.25}$, and the spiral arm regions have $\log_{10}(n_{\rm 0,H_2})=3^{+0.25}_{-0.25}$. Finally, the density dispersion is wider in the center ($\sigma=0.45^{+0.35}_{-0.25}$) relative to the bar ends ($\sigma=0.25^{+0.05}_{-0.05}$) and spiral arms ($\sigma=0.3^{+0.05}_{-0.10}$). We note that our parameter space is limited to the extent covered within the defined grid. Therefore, values outside of our coverage are highlighted in red in \autoref{fig:model_results}.

Given our consistent finding that the mean density of the CO-emitting gas is around $n_{0, \rm H_2}=10^3\,$cm$^{-3}$, we qualitatively investigate the correlation of the CO line ratios again in \autoref{fig:model_grid} by comparing the distribution of modeled and observed line ratios. To first order, we find that keeping the mean volume density, $n_{0, \rm H_2}$, and density dispersion, $\sigma$ fixed, the overall positive trend is well characterized by increasing temperature. The $R_{21}$ ratio seems to be more sensitive to temperature than $R_{31}$, particularly above 30\,K and at low $N_{\rm CO}/\Delta v<10^{16.5}\,\rm cm^{-2}\,km^{-1}\,s$. Looking at the temperature lines in the figure, we see that a fixed temperature line covers a smaller $R_{21}$ range than $R_{31}$. {{Therefore}}, the scatter {{along the correlation}}, which we see in \autoref{fig:model_grid}, is the result of varying $N_{\rm CO}/\Delta v$, which traces changes in the optical depth of the emission. {For completeness,  we report that with respect to the other ratio-to-ratio comparisons (see \autoref{fig:ratio_vs_ratio}), we find also for the $R_{32}$-to-$R_{21}$ correlation that the scatter is mainly driven by $N_{\rm CO}/\Delta v$ variation. For the $R_{32}$-to-$R_{31}$ relation which shows the smallest scatter ($\sigma=0.062\,\rm dex$), we do not identify a single driver for the scatter as both temperature and $N_{\rm CO}/\Delta v$ equally affect the scatter.}

\subsubsection{Interpreting CO line ratio variations}
\label{disc:overall}
Combining our observed correlations of galaxy properties with the CO line ratio variation and the derived molecular gas conditions from the line modeling, we focus this discussion on the three environments of interest in NGC\,3627.

\noindent {\bf Center:} The center has CO line ratios that are elevated by 30-50\% relative to the galaxy-wide average. This ``hook" of the ratios toward lower radii is constrained to the inner $<1\,{\rm kpc}$ in radius. Therefore, high angular resolution observations are required to detect this increase. At coarser, kpc-scale observations, this ''hook" will flatten out and appear as a more gradual radial trend (see \autoref{fig:varyres}). Our line modeling approach suggests that this is connected to higher temperatures and a wider density width. Contrasting the line ratio variation to the SFR (left panels in \autoref{fig:ratio_comp}) we see that the center has a clearly offset trend toward lower SFRs relative to the other regions, in accordance with suppressed star formation in this region. This could be connected to the presence of an AGN in the center of NGC\,3627, which may quench or lower star formation \citep[e.g.,][]{Birnboim2003,Gabor2010}.  Therefore the higher temperatures may then be related to heating caused by the AGN. 

Furthermore, we note that other studies report even higher kinetic temperatures for the center of NGC\,3627. For instance, using single-zone \texttt{RADEX} models, \cite{Teng2023} derive temperatures of $T_{\rm kin}{>}100\,$K in the central region. Similarly, \citet{Teng2022, Teng2023} and \citet{Liu2023} find $T_{\rm kin}{>}100\,$K also in other nearby galaxy centers. If only the low-$J$ \chem{^{12}CO} ratios are used to constrain $T_{\rm kin}$, as presented in \autoref{fig:model_results}, then we require $R_{21}\ge1$ to get $T_{\rm kin} > 50$ K. However, because  $^{12}$CO(2-1) has a transition energy of $E_u$ of 16.6 K, it is easily thermalized. Hence, in principle $R_{21}$ is insensitive of $T_{\rm kin}$ unless the volume density is much lower than the critical density. The same reasoning holds for $R_{32}$, with CO(3-2) having a slightly higher transition energy, but still below $100$\,K. This also explains why with a temperature of ${\sim}25\,$K, \citet{Law2018} find a similar temperature to our result, as they rely only on the low-$J$ CO ratio as well.

\medskip 

\noindent{\bf Bar-Ends:} The bar ends have higher line ratios than the galaxy-wide average. But the measured mean temperature, volume density, and density dispersion are similar to those in the disk. However, while the mean is similar, the temperature distribution shows a stronger tail towards higher $T_{\rm kin}$, up to 25\,K within the 16$^{\rm th}$-84$^{\rm th}$ percentile range. The higher ratios, therefore, are likely due to heating caused by the increased SFR in this region \citep[e.g.][]{Warren2010, Casasola2011}. This is further supported by the consistent CO trends with SFR and the IR dust tracers for the bar, bar ends, and spiral arm, where the primary difference is a higher average SFR in the bar ends. Such high SFRs could result from enhanced cloud-cloud collisions that are predicted for the bar ends \citep[e.g.][]{Takahira2014,Maeda2021}. But our derived density dispersion does not likely support cloud-cloud collisions, as we would expect a flatter density tail for this region. We note,  however, that using only the $^{12}$CO line ratios, we could also just not be sensitive to variation in the density, and hence not detect such a change. To fully rule out cloud-cloud collisions, we require the observation of higher-density tracing molecular lines that are expected to be more sensitive to changes in the density dispersion \citep[e.g.][]{Leroy2017_density}. 
\medskip 

\noindent{\bf Overall:}
The positive correlation we find for all three line ratio comparisons (see the three panels in \autoref{fig:ratio_vs_ratio}) indicates that the higher-$J$ transitions are more affected by changes in the excitation conditions than variations in other physical parameters. Hotter gas, at least within the range of values we expect to find across nearby spiral galaxies ($T{\sim}10{-}50\,$K) will drive up the level populations of the higher energy levels more significantly. Contrasting the critical density of the $J{=}3{\rightarrow}2$ and $J{=}1{\rightarrow}0$ transitions, their energy difference amounts to ${\sim}27.7$\,K and the critical density of the latter is an order of magnitude lower \citep{Penaloza2018}. This explains why the $J{=}3{\rightarrow}2$ increases so significantly with respect to the lower-$J$ transition in the hotter and/or denser center and bar-ends of the galaxy (the visible ``hook" observed in the bottom two panels of \autoref{fig:ratio_vs_ratio} where $R_{21}$ is on the $y$-axis). 
Our findings suggest therefore that the CO line ratio yields a diagnostic for tracing changes in  the temperature of the molecular gas. This strong connection to the heating sources across the galaxy makes it possible to use tracers, such as $\Sigma_{\rm SFR}$ or the JWST IR bands to parametrize changes in the ratios even at high angular resolutions. Only in the center did we see that we need to adjust the relation (i.e. we measure a systematic offset of the trend line), likely due to additional heating by the AGN.  However, the ratios do not seem to be sensitive to variations in volume density. To trace density variations more effectively, we likely require observations of lines with higher critical density, such as for example, HCN, HNC, or N$_{2}$H$^+$. 
\section{Conclusions}
\label{sec:conclusion}
In this study, we investigate the CO excitation via the integrated CO line ratios $R_{21}$, $R_{32}$, and $R_{31}$ across center, bar, and spiral arm of NGC\,2903 and NGC\,3627. We compare the distribution of line ratios across the distinct regions of the galaxy and link them to the physical conditions of the molecular gas (such as temperature and density).

\begin{enumerate}
    \item In NGC\,3627, the center shows higher line ratios in the central ($R_{\rm gal}{<}0.5$\,kpc) region compared to the average disk value by ${\sim}35\%$ ($R_{21}$), ${\sim}50\%$ ($R_{32}$), and ${\sim}66\%$ ($R_{31}$).
    \item  We compare the trend of $R_{32}$ in NGC\,3627 to NGC\,2903, where we find an identical pattern. Also NGC\,2903 shows a steep increase of the CO line ratio toward the center at $r{<}1\,$kpc. This highlights that the center of galaxies have extreme environments that can significantly affect the CO SLED.
    \item In NGC\,3627, we measure a consistent trend of the line ratios with star formation rate surface density, $\Sigma_{\rm SFR}$ across the different environments. The slope of the relation in log-log space is $m{\approx}0.1$ for  $R_{21}$, $m{\approx}0.2$  for $R_{32}$, and $m{\approx}0.3$ for $R_{31}$.
    \item We also identify a positive correlation with the mid-IR 7.7\,$\mu$m and 21\,$\mu$m narrow band emission. The 7.7\,$\mu$m correlation is steeper than the 21\,$\mu$m for all three line ratios This correlation reflects that higher line ratios probe regions of hotter kinetic gas temperature, potentially due to stronger interstellar radiation fields.
    \item The distribution of line ratios across the galaxy differs depending on the resolution of the observed data. We find systematically lower average values at 28$''$ than at $4''$ by ${\sim}0.1$\,dex.
\end{enumerate}

We compare the observed and measured line ratios across the different environments to predicted line ratios from \texttt{RADEX} modeling.

\begin{enumerate}[resume]
    \item We perform a $\chi^2$ minimization for the individual sightlines. While we find significant differences in the line ratio distribution between the center and arm-bar region of NGC\,3627, we do not identify significant differences in the maximum likelihood of the distribution of physical parameters, such as the mean volume density, $n_{0, \rm H_2}$, or lognormal density dispersion, $\sigma$. Only our derived temperatures are higher, on average, by 10-15\,K in the center than the spiral arms.
\end{enumerate}

Overall, our results suggest that the low-$J$ CO line ratios are sensitive to chances in temperature of the bulk molecular gas. This is also likely an explanation the tight correlation of the line ratios with $\Sigma_{\rm SFR}$ and the IR-based dust tracers, which scale with the heating sources.  In contrast, the CO line ratios do not seem to effectively trace changes in the underlying volume density of the CO emitting medium.



\section*{Acknowledgements}
    This work was carried out as part of the PHANGS collaboration. {We would like to thank the referee for going carefully through the paper and appreciate the constructive comments that improved this paper.}
      JdB and EWK acknowledge support from the Smithsonian Institution as a Submillimeter Array (SMA) Fellow. 
      The work of AKL is partially supported by the National Science Foundation under Grants No. 1615105, 1615109, and 1653300.
      AU acknowledges support from the Spanish grant PID2019-108765GB-I00, funded by MCIN/AEI/10.13039/501100011033. 
      ES acknowledges funding from the European Research Council (ERC) under the European Union’s Horizon 2020 research and innovation programme (grant agreement No. 694343).
      ER acknowledges the support of the Natural Sciences and Engineering Research Council of Canada (NSERC), funding reference number RGPIN-2022-03499.
      MQ and MJJD acknowledge support from the Spanish grant PID2019-106027GA-C44, funded by MCIN/AEI/10.13039/501100011033.
      MC gratefully acknowledges funding from the Deutsche Forschungsgemeinschaft (DFG) through an Emmy Noether Research Group (grant number CH2137/1-1). COOL Research DAO is a Decentralized Autonomous Organization supporting research in astrophysics aimed at uncovering our cosmic origins. SCOG acknowledges support from the DFG via SFB 881 “The Milky Way System” (sub-projects B1, B2 and B8) and from the Heidelberg cluster of excellence EXC 2181-390900948 “STRUCTURES: A unifying approach to emergent phenomena in the physical world, mathematics, and complex data”, funded by the German Excellence Strategy. He also acknowledges support from the European Research Council in the ERC synergy grant ‘ECOGAL’ Understanding our Galactic ecosystem: From the disk of the Milky Way to the formation sites of stars and planets’ (project ID 855130).
      Y-HT acknowledges funding support from NRAO Student Observing Support Grant SOSPADA-012 and from the National Science Foundation (NSF) under grant No. 2108081.
      TGW acknowledges funding from the European Research Council (ERC) under the European Union’s Horizon 2020 research and innovation programme (grant agreement No. 694343). 

\section*{Data Availability}
The data underlying this article will be shared on reasonable request to the corresponding author.


\bibliographystyle{mnras}
\bibliography{references.bib} 




\appendix
\section{The CO(3-2) ACA Data of NGC\,2903 and 3627}
\begin{figure*}
    \centering
    \includegraphics[width=0.7\textwidth]{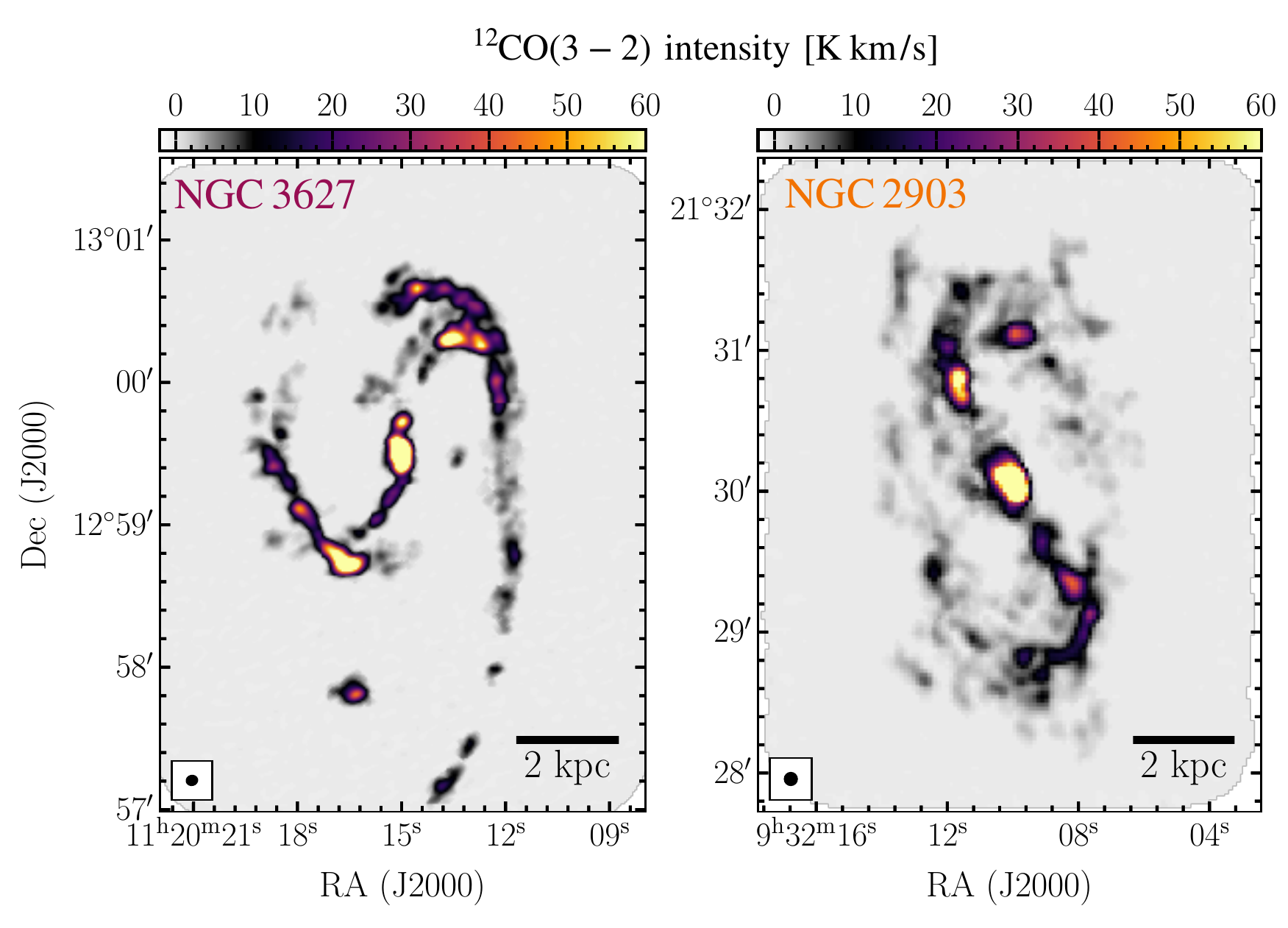}
    \caption{{\bf ACA 7m \chem{^{12}CO}{32} moment-0 maps of NGC\,3627 (left) and NGC\,2903 (right).} The moment-0 map is computed by integrating over channels that contain emission. These are determined by growing a high into a low S/N cut ($\sigma{>}6$ and $\sigma{>}3$, respectively). To illustrate the extent of the map, we added a signal-free channel to the map.  }
    \label{fig:co32_mom0map}
\end{figure*}

In this study, we present new ACA 7m moment-0 maps of NGC\,2903 and NGC\,3627 (PI: J. Puschnig). These maps have not yet been presented in previous studies. For completeness, we present here the moment-0 maps showing the entire extent of the observations. For the line ratio analysis, we only include the field of view where all CO line maps overlap. In \autoref{fig:co32_mom0map} we present the moment-0 map for full field of view of the \chem{^{12}CO}{32} observation. Note that for NGC\,3627, this CO(3-2) dataset also includes the full extent of the southern spiral arm (which we do not include in our analysis as we lack \chem{^{12}CO}{10} coverage).

\section{Variation of the CO at different angular resolutions}
\label{sec:resol_analysis}

\begin{figure}
    \centering
    \includegraphics[width = \columnwidth]{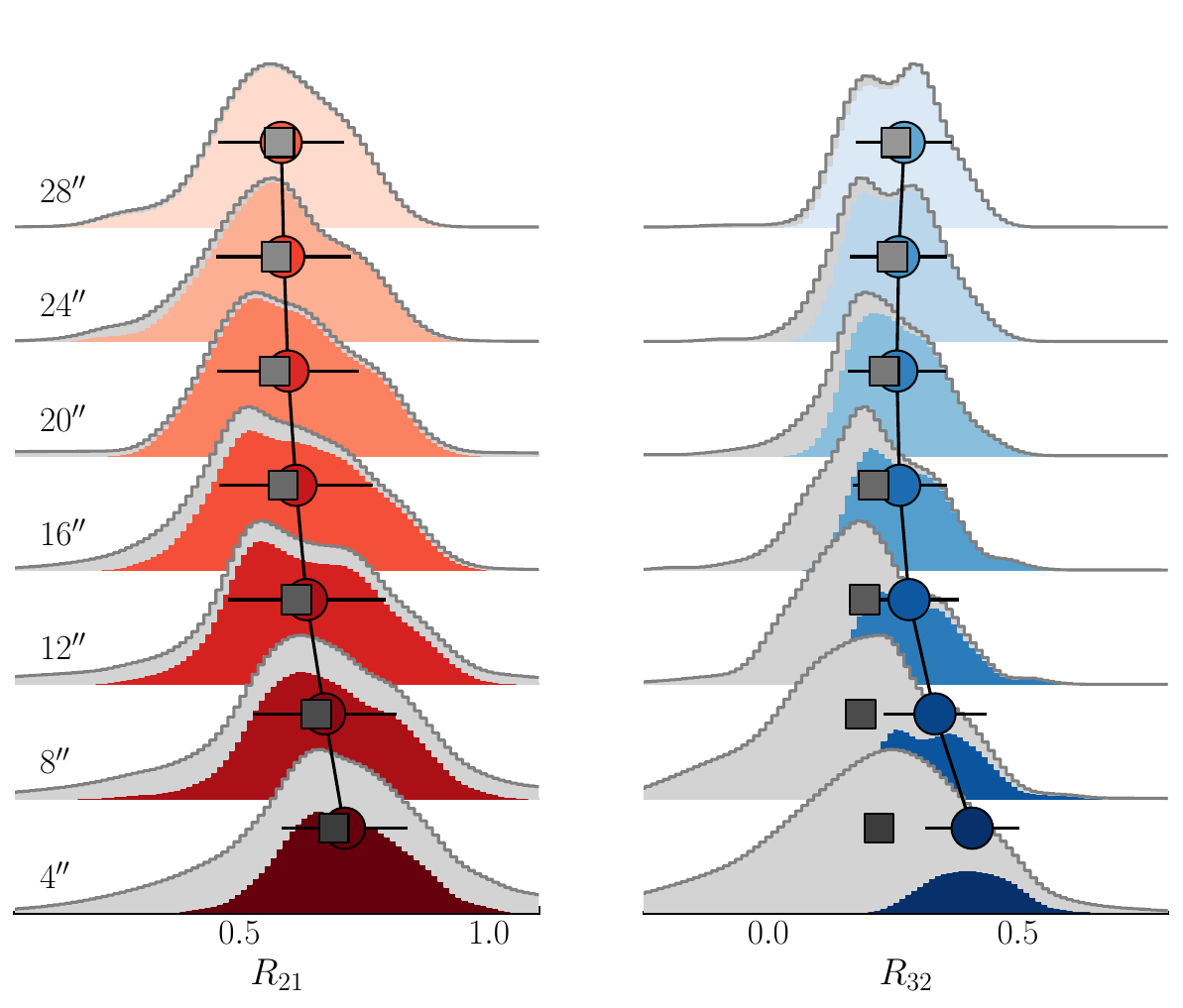}
    \caption{{\bf Peak Normalized CO Line Ratio Distribution with Resolution} We convolved the CO data cubes to different angular resolutions from 4$''$ to $28''$. Significant CO line ratios are indicated in color. The circle shows the median line ratio only including the significant data points, while the grey square indicated the sample wide median including all sightlines. The horizontal black line indicates the 1$\sigma$ scatter for the significant CO data points.}
    \label{fig:ratio_res}
\end{figure}

In \autoref{fig:ratio_res}, we compare the distribution of the CO line ratios after convolving our CO data cubes to increasingly lower angular resolutions from 4$''$ to $28''$ in steps of $\Delta\theta{=}4''$. An angular resolution of $28''$ corresponds to the angular resolution of the data analyzed for the $R_{21}$ line ratio in \citet{denbrok2021}. This analysis hence demonstrates whether the distribution in $R_{21}$ and $R_{32}$ varies between kpc-scale (${\ge}20''$) and cloud-scale (${\le}4''$) studies. We find that the mean of the distribution of both $R_{21}$ and $R_{32}$ is higher at the higher angular resolution. Qualitatively, this can be seen in \autoref{fig:ratio_res}, as the mean point moves towards lower $R_{21}$ for coarser angular resolutions. At $4''$, we have $\langle R_{21}^{4''}\rangle = 0.72\pm0.12$ for the full distribution, which decreases to $\langle R_{21}^{28''}\rangle = 0.62\pm0.11$. We find the same effect for $R_{32}$, where at higher resolutions, $\langle R_{32}^{4''}\rangle = 0.4\pm0.10$, while we find a much lower value at coarser resolutions of  $\langle R_{32}^{28''}\rangle = 0.27\pm0.09$. \autoref{tab:res_ratio_resolution} lists the mean and standard deviations for the CO line ratios for the different resolutions. Performing a Kolmogornov-Smirnov test \citep{Smirnov1939} by comparing the coarser resolution ones to the distribution at $4''$, we find that all distributions differ significantly. We note that also when accounting for upper and lower limits to our line ratios as censored data with the \texttt{logrank\_test} function from the Python package \texttt{lifelines}, we find that all comparisons show $p{\ll}0.05$.  Such a difference in the overall distribution has consequences when adopting line ratios from previous studies that use observations at different angular resolutions.

Such a trend in the mean of the $R_{21}$ CO line ratio at different angular resolutions is also reported in other nearby galaxies (NGC\,0628, NGC\,4254; T. Saito et al. in prep). A coarser beam will incorporate a larger variety of different environments. The observation will mix emission from different clouds and inter-cloud regions. In particular, the latter will likely lower the resulting CO line ratio value due to lower excitation conditions in such environments. We note that due to the improvement of the spectral rms after convolution, we could bias our result towards lower $R_{21}$ since, due to the improved S/N, we would detect emission from a more extended region. We assess that this effect, however, cannot explain the decreasing trend. The fact that after convolution, the CO line ratios, on average, decrease indicates that the overall contrast of the higher-$J$ emission is stronger than for the $J=1\rightarrow0$ emission. Hence, we conclude that at high resolution, we can resolve individual regions of molecular gas at high densities or where the interstellar radiation field locally increases the excitation that contributes towards high CO line ratios. At coarser resolutions, these environments are washed out due to beam smearing.

\begin{table*}
    \centering
    \begin{threeparttable}
    
    \caption{CO Line Ratio at Different Resolution Summary}
    \label{tab:res_ratio_resolution}
    
    \begin{tabular}{l c c c c | c c c c}\hline
          & \multicolumn{4}{c}{$R_{21}$} & \multicolumn{4}{c}{$R_{32}$}  \\
          & $\mu$ & $s$ & $n$ & $p$ (KS) &$\mu$ & $s$ & $n$ &  $p$ (KS)\\\hline\hline
         $4''$  & 0.72 & 0.12 & 1868 & $-$              &   0.40 & 0.10 & 596 & $-$ \\
         $8''$  & 0.68 & 0.14 & 762  & $5.5\times 10^{-8}$  &   0.31 & 0.10 & 398 & $6.7\times10^{-26}$\\
         $12''$ & 0.65 & 0.15 & 439  & $2.0\times 10^{-15}$ &   0.28 & 0.10 & 287 & $3.1\times10^{-38}$\\
         $16''$ & 0.63 & 0.15 & 272  & $8.9\times 10^{-16}$ &   0.26 & 0.09 & 239 & $1.4\times10^{-46}$ \\
         $20''$ & 0.62 & 0.13 & 188  & $4.6\times 10^{-17}$ &   0.25 & 0.09 & 195 & $7.6\times 10^{-47}$\\
         $24''$ & 0.60 & 0.12 & 133  & $4.4\times 10^{-16}$ &   0.26 & 0.10 & 153 & $6.6\times 10^{-41}$\\
         $28''$ & 0.61 & 0.11 & 108  & $2.6\times 10^{-13}$ &   0.27 & 0.09 & 121 & $3.3\times 10^{-16}$\\\hline
    \end{tabular}
 
    \begin{tablenotes}
      \small 
      \item {\bf Notes:} The table lists the mean ($\mu$), standard deviation ($s$) and number of data points ($n$) per distribution. The significance $p$ indicates whether the distribution differs from the distribution at $4''$ resolution based on a Kolmogorov-Smirnov test \citep{Smirnov1939}.
    \end{tablenotes}
    \end{threeparttable}
    
\end{table*}

\section{Line Modeling}
\label{app_model}
\begin{figure*}
    \centering
    \includegraphics[width = \textwidth]{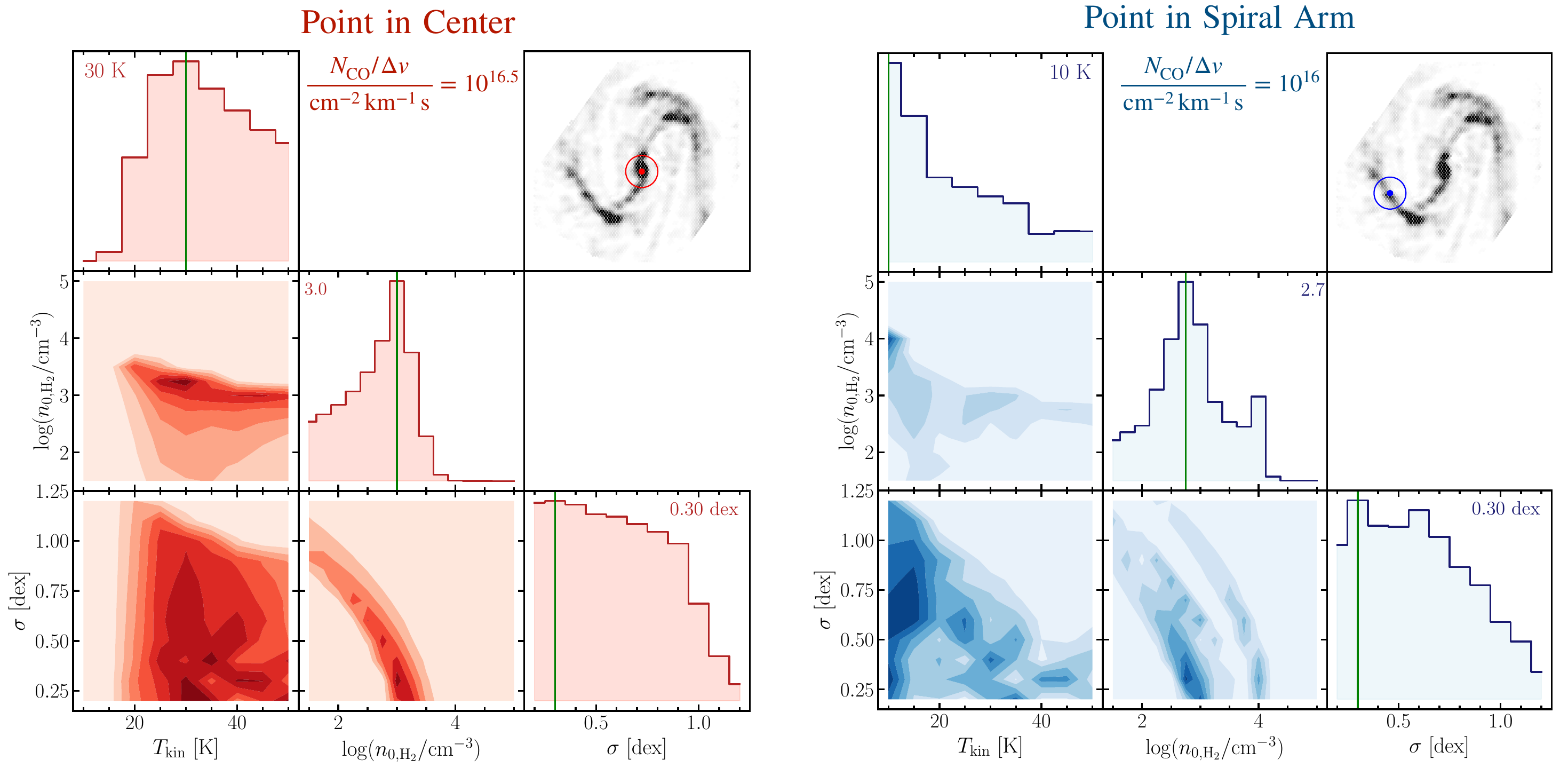}
    \caption{{\bf Constraining the Physical Conditions using a $\chi^2$ minimization} The corner plot shows the 1D and 2D PDF for the three different parameters. The heat map is the 2D PDF for the different permutations of the parameters. Left panel (in red) is for an arbitrary point the \emph{Center} region and right (in blue) for an arbitrary point in the \emph{Spiral Arm}. The top right panel shows a map of NGC\,3627 in \chem{^{12}CO}{21} emission, where the overlaid circle indicated in red/blue is the sightline for which the PDF of the parameters is computed. Each 1D histogram indicates the maximum likelihood value.}
    \label{fig:chi2_min}
\end{figure*}

To match the observed and modeled line ratios, we perform a $\chi^2$ minimization for each line of sight. \hyperref[fig:chi2_min]{Figure~\ref*{fig:chi2_min}} illustrates the 1D and 2D  likelihood distributions for a representative sightline within the central and spiral arm regions for the set of three physical parameters. In the example given, the central region shows high temperatures with $T_{\rm kin}{\ge} 30$\,K. For our analysis, we fix the column density per line width for each line of sight. The maximum likelihood determines the value we assign to each sightline. 


\bsp	
\label{lastpage}
\end{document}